\documentclass[sigconf]{acmart}

\AtBeginDocument{%
  }

\copyrightyear{2025}
\acmYear{2025}
\setcopyright{cc}
\setcctype{by}
\acmConference[CHI '25]{CHI Conference on Human Factors in Computing Systems}{April 26-May 1, 2025}{Yokohama, Japan}
\acmBooktitle{CHI Conference on Human Factors in Computing Systems (CHI'25), April 26-May 1, 2025, Yokohama, Japan}\acmDOI{10.1145/3706598.3714132} \acmISBN{979-8-4007-1394-1/25/04}

\usepackage{listings}
\usepackage{subcaption}
\usepackage{float}

\lstdefinestyle{mystyle}{
    basicstyle=\ttfamily\footnotesize
}

\begin{document}

\title{Tactile Vega-Lite: Rapidly Prototyping Tactile Charts with Smart Defaults}

\author{Mengzhu (Katie) Chen}
\orcid{0009-0001-6404-7647}
\affiliation{%
  \institution{CSAIL\\Massachusetts Institute of Technology}
  \city{Cambridge}
  \state{Massachusetts}
  \country{USA}
}
\email{mzc219@mit.edu}

\author{Isabella Pedraza Pineros}
\orcid{0009-0002-3269-7618}
\affiliation{%
  \institution{CSAIL\\Massachusetts Institute of Technology}
  \city{Cambridge}
  \state{Massachusetts}
  \country{USA}
}
\email{ipedraza@mit.edu}

\author{Arvind Satyanarayan}
\orcid{0000-0001-5564-635X}
\affiliation{%
  \institution{CSAIL\\Massachusetts Institute of Technology}
  \city{Cambridge}
  \state{Massachusetts}
  \country{USA}
}
\email{arvindsatya@mit.edu}

\author{Jonathan Zong}
\orcid{0000-0003-4811-4624}
\affiliation{%
  \institution{CSAIL\\Massachusetts Institute of Technology}
  \city{Cambridge}
  \state{Massachusetts}
  \country{USA}
}
\email{jzong@mit.edu}

\renewcommand{\shortauthors}{Chen et al.}

\begin{abstract}
Tactile charts are essential for conveying data to blind and low vision (BLV) readers but are difficult for designers to construct. Non-expert designers face barriers to entry due to complex guidelines, while experts struggle with fragmented and time-consuming workflows that involve extensive customization. Inspired by formative interviews with expert tactile graphics designers, we created Tactile Vega-Lite (TVL): an extension of Vega-Lite that offers tactile-specific abstractions and synthesizes existing guidelines into a series of smart defaults. Predefined stylistic choices enable non-experts to produce guideline-compliant tactile charts quickly. Expert users can override defaults to tailor customizations for their intended audience. In a user study with 12 tactile graphics creators, we show that Tactile Vega-Lite enhances flexibility and consistency by automating tasks like adjusting spacing and translating braille while accelerating iterations through pre-defined textures and line styles. Through expert critique, we also learn more about tactile chart design best practices and design decisions. 
\end{abstract}



\maketitle

\section{Introduction}
Tactile charts are an essential tool for blind and low vision (BLV) people to independently explore data and participate equally in discussions involving statistical analysis~\cite{engelUserStudyDetailed2019, watanabeDevelopmentSoftwareAutomatic2012}. As a subset of tactile graphics, tactile charts use braille, raised lines, textured areas, and distinct marks to represent statistical data through touch \cite{engelImproveAccessibilityTactile2017}. Research has found that tactile charts facilitate independent data exploration, communicate spatial relationships, provide non-sequential access to information, and support content comprehension and memorization \cite{engelUserStudyDetailed2019}.
They surpass printed or electronic tables in helping blind users understand the correlations between variables~\cite{watanabeDevelopmentSoftwareAutomatic2012}, helping students learn statistical concepts, and developing tactile graphic literacy \cite{engelImproveAccessibilityTactile2017, aldrichFirstStepsModel2003}. 

However, when we conducted formative interviews with tactile graphics designers (\autoref{sec:motivation}), we found that the existing workflows for designing and prototyping tactile charts are, unfortunately, tedious and time-consuming, requiring a high level of skill and experience.
We found that because best practices for tactile design differ significantly from practices in visual design, non-expert designers face significant barriers to entry. For example, in visual design, it is common to optimize for space efficiency or aesthetics by orienting text sideways or scaling it while still maintaining legibility. However, this practice does not translate well to tactile design. Braille, the standard form of text in tactile charts, cannot be scaled, and orienting it in different directions can make it impossible to read. Prioritizing space efficiency can lead to overcrowding, which hinders the clarity and usability of tactile charts, especially for non-expert designers. Existing guidelines for tactile graphic design have a steep learning curve and can be overwhelming for those new to the field. 
Even for expert tactile designers, difficulties arise due to a lack of dedicated software tools for creating tactile charts.
Designers typically need to master multiple tools to create a single chart, including general-purpose vector drawing tools and braille translation software.
Each has its own learning curve, and designers must repeat tedious operations, such as transferring designs between tools, to make manual adjustments, even for minor modifications.
The labor-intensive, specialized work involved in this process hinders the adoption of tactile charts in practice.

To address these challenges, we present Tactile Vega-Lite (TVL): extensions to Vega-Lite \cite{satyanarayanVegaLiteGrammarInteractive2017} for rapidly prototyping tactile charts.
In contrast to existing workflows that rely on fragmented functionality across various general-purpose tools, TVL introduces a set of domain-specific abstractions for tactile chart design (\autoref{sec:system-design}).
These include support for tactile encoding channels, braille, navigational aids, and layout configuration.
These extensions to Vega-Lite enable designers to benefit from the advantages of existing visualization grammars---such as the ability to systematically enumerate a design space---while still expressing tactile-specific affordances.
They also enable authors of Vega-Lite visualizations to make small edits to existing visual specifications to create tactile charts without requiring deep tactile design expertise.

Just as Vega-Lite provides \textit{smart defaults} that enable authors to easily specify common visual chart forms, TVL's defaults generate tactile charts that align with established guidelines.
For example, TVL renders grid lines as the least tactually distinct lines on a graph, as recommended by the Braille Authority of North America (BANA) guidelines \cite{brailleauthorityofnorthamericaGuidelinesStandardsTactile2022}. 
Also, TVL automatically places the x-axis title below the labels and aligns it to the leftmost x-axis label, where readers conventionally expect to find it.
Expert designers can override these defaults to make context-specific adjustments for their output medium (e.g. braille embosser, swell paper) or intended audience.
Similarly, predefined palettes of lines and area textures allow non-experts to choose from a set of reliable options while enabling experts to rapidly prototype with reusable texture configurations.

We implement a prototype TVL editor that takes a declarative TVL specification and outputs a tactile chart in SVG format.
These SVGs can be printed using various production methods; popular options include an embosser or a swell form machine.
Through an example gallery (\autoref{sec:example-gallery}), we demonstrate that TVL expresses a variety of chart forms and layouts.
We also conducted a user study (\autoref{sec:user-study}) with 12 tactile graphics designers to understand how TVL compares to their existing design workflows.
We found that participants valued predefined textures and line styles for consistency and efficiency, aligning with existing workflows. Customizability was crucial for adapting tactile charts, especially for younger readers who benefit from grid lines and clear spacing. Differences in expectations between professional designers and Teachers of Students with Visual Impairments (TVIs)\footnote{
The term ``Teacher of Students with Visual Impairments'' (TVI) is widely used by organizations such as the American Printing House for the Blind (APH) and the Braille Authority of North America (BANA) to refer to educators who work with blind and low vision students. 
We acknowledge that the term ``visually impaired'' can be considered harmful language \cite{athersharifAmDisabledNot2024}, or otherwise not preferred \cite{hansonWritingAccessibility2015}.
In this paper, we avoid using this term to describe people.
However, we use TVI for clarity when referencing the specific job title, to maintain consistency with its use by educational institutions like APH \cite{garcia-mejiaCentralRoleTeacher2024}, by educators \cite{carmenwillingsTeacherStudentsVisual2020}, and by the state in licensing teachers \cite{anthonyDefinitionRoleTVI2005}.}
 revealed a trade-off between advanced design capabilities and practical, easy-to-use solutions, with professionals seeking more complex tools and TVIs favoring simplicity and efficiency.
Through an expert critique of the system's defaults, we found that designers appreciated the tactile-specific design assets, such as predefined textures and line styles, which encouraged reasoning and prototyping to determine the optimal design for different audiences. This feedback validates our dual approach of providing guideline-aligned defaults while allowing for customization to accommodate context-specific needs, balancing efficiency, functionality, and reader experience.

\section{Related Works}
\subsection{Tactile Chart Design}

The effectiveness and legibility of tactile charts are heavily influenced by their design, particularly how well the chart accommodates the tactile reader’s ability to discern shapes, lines, and textures through touch. To inform design, researchers have studied the relative effectiveness of different types of tactile charts, such as diagrams \cite{goncuTactileDiagramsWorth2010}, heatmaps \cite{engelTactileHeatmapsNovel2021}, schematics \cite{raceDesigningTactileSchematics2019}, and network graphs \cite{yangTactilePresentationNetwork2020}.
Another body of work studies the effectiveness of specific tactile design elements, such as textures \cite{prescherConsistencyTactilePattern2017, watanabeEffectivenessTactileScatter2018, purdueuniversityTactileDiagramManual2002, watanabeTexturesSuitableTactile2018}, or grid lines \cite{doi:10.1177/001872088402600106, ledermanTangibleGraphsBlind1982, aldrichTangibleLineGraphs1987}. These studies highlight the need for deliberate design choices that prioritize tactile discernibility.

The reader's experience with tactile reading also affects the effectiveness of a particular design. Experienced tactile readers, for example, can more easily interpret dense information and advanced tactile symbols, while novice readers may require simpler charts with fewer details. Tactile reading strategies can vary, with some individuals using the palm of their hand to explore larger areas and their fingertips for more detailed inspection \cite{hospitalTipsReadingTactile2024, hastyTeachingTactileGraphics2024}. As a result, the design of the tactile chart must be tailored to its intended audience.

Finally, the methods and materials with which a tactile chart is produced can affect a reader's experience.
The most commonly used production methods for creating tactile charts include microcapsule paper, embossers, or vacuum-forming techniques.
Tactile charts can also be viewed as a form of data physicalization \cite{jansenOpportunitiesChallengesData2015}, where data is transformed into physical artifacts to leverage human tactile perception.
For instance, the use of collages and 3D models has been less common due to their limited replicability but is popular among teachers of blind and low-vision students \cite{prescherProductionAccessibleTactile2014}. 
Refreshable braille displays \cite{hollowayRefreshableTactileDisplays2024} provide digital tactile output, enabling users to access dynamic content; however, they are limited in size and struggle to represent detailed graphics. The choice of production method matters because each method offers a different tactile experience and has implications for the chart's longevity, texture fidelity, and ease of reading.

In our work, we create an authoring system for tactile charts intended for use with embossers or swell form paper.
We incorporate best practices for tactile chart design into our system's default specifications, making it easy for an author to rapidly create an effective chart.
An author can make small edits to a declarative specification to make adjustments for their intended audience.

\subsection{Authoring Tactile Charts}
The most common way to author tactile charts is through manual authoring using vector drawing tools, such as Adobe Illustrator and Corel Draw, or physical materials like braille graph paper and raised-line drawing kits. These methods offer high control over the final tactile output but are labor-intensive and require specialized knowledge of tactile design principles. A designer must meticulously adjust line thickness, spacing, and texture to ensure a chart is legible and informative for tactile users. This often involves trial and error, with multiple revisions needed to achieve the desired outcome.

Automated and semi-automated systems for authoring tactile charts provide an alternative to manual creation, offering more efficient workflows. These systems are generally either image-driven systems or data-driven systems. Image-driven systems, such as those using image processing, automatically convert existing visual charts into tactile formats \cite{wayAutomaticVisualTactile1997, mechEdutactileToolRapid2014, krufkaVisualTactileConversion2007a, crombieBiggerPictureAutomated2004, jayantAutomatedTactileGraphics2007, 10.1145/3640543.3645175, hollowayRefreshableTactileDisplays2024, gonzalezTactiledMoreBetter2019}. While these methods are effective for general image conversion, they are less ideal for visualizations. The main issues include a loss of data granularity and a compromise in fidelity, which can obscure critical information.  
Data-driven systems, by contrast, directly convert data into tactile charts without relying on a visual reference.
One such example is SVGPlott \cite{engelSVGPlottAccessibleTool2019}, a tool that transforms structured data (such as CSV files) into tactile visualizations.
However, existing systems exhibit several shortcomings in expressiveness and customizability \cite{watanabeDevelopmentSoftwareAutomatic2012, watanabeDevelopmentTactileGraph2016, engelSVGPlottAccessibleTool2019, goncuTactileChartGeneration2008}. These include restrictive output formats (PNG) that limit post-creation editing \cite{watanabeDevelopmentTactileGraph2016}, a lack of support for custom visual encodings \cite{engelSVGPlottAccessibleTool2019}, and inadequate compliance with established accessibility guidelines \cite{watanabeDevelopmentSoftwareAutomatic2012, goncuTactileChartGeneration2008}. Additionally, these systems do not support advanced data transformations or iterative design processes, which are essential for refining and optimizing tactile charts. The limited customization options, especially regarding textural enhancements and detailed layout adjustments, further constrain their effectiveness and usability.

In our work, we create an automated, data-driven authoring system. Our work aims to strike a balance between rapid authoring through guideline-compliant defaults and customizability through tactile abstractions for encoding, layout, navigational aids, and braille. 
\section{Motivation}
\label{sec:motivation}

To better understand the practical challenges involved with making tactile charts, we conducted formative interviews with tactile graphics experts to motivate our system design. From these interviews, we synthesized a set of design challenges that inform the design of tactile chart authoring systems.
In this section, we first describe our process for conducting formative interviews. Then, we summarize a tactile chart designer's workflow: first by walking through an example of a professional designer's workflow and then discussing how an educator's workflow might differ.
Finally, we discuss how these workflows translate into design challenges that tactile chart authoring systems can address.

\subsection{Methods}
We conducted seven formative interviews with experts in the field, including three Teachers of Students with Visual Impairments (TVIs), one assistive technology specialist who creates tactile graphics, two developers of assistive technologies, and one professional tactile graphics designer. Experts were sourced through snowball sampling, leveraging connections through acquaintances, and reaching out to relevant institutes and organizations, such as the Andrew Heiskell Braille and Talking Book Library in New York City. The interviews, lasting between 30 minutes to 1 hour, were conducted both in person and online. We asked participants about the current state of tactile graphics, their perception of existing tools, best practices in tactile graphics design and ways to learn about it, how they create tactile graphics, and their production methods.

Additionally, we conducted an hour-long observational session with a professional tactile graphics designer (one of our seven interview participants) to better understand their design and authoring workflow. During the session, we observed the designer creating new charts and updating and adjusting existing charts in Adobe Illustrator. The designer explained their process and reasoning aloud as they worked, giving us valuable insights into their decision-making approach.

\subsection{Expert Tactile Chart Designer Workflows}
\label{sec:designer-workflows}
\begin{figure*}[!ht]
    \includegraphics[width=0.49 \linewidth]{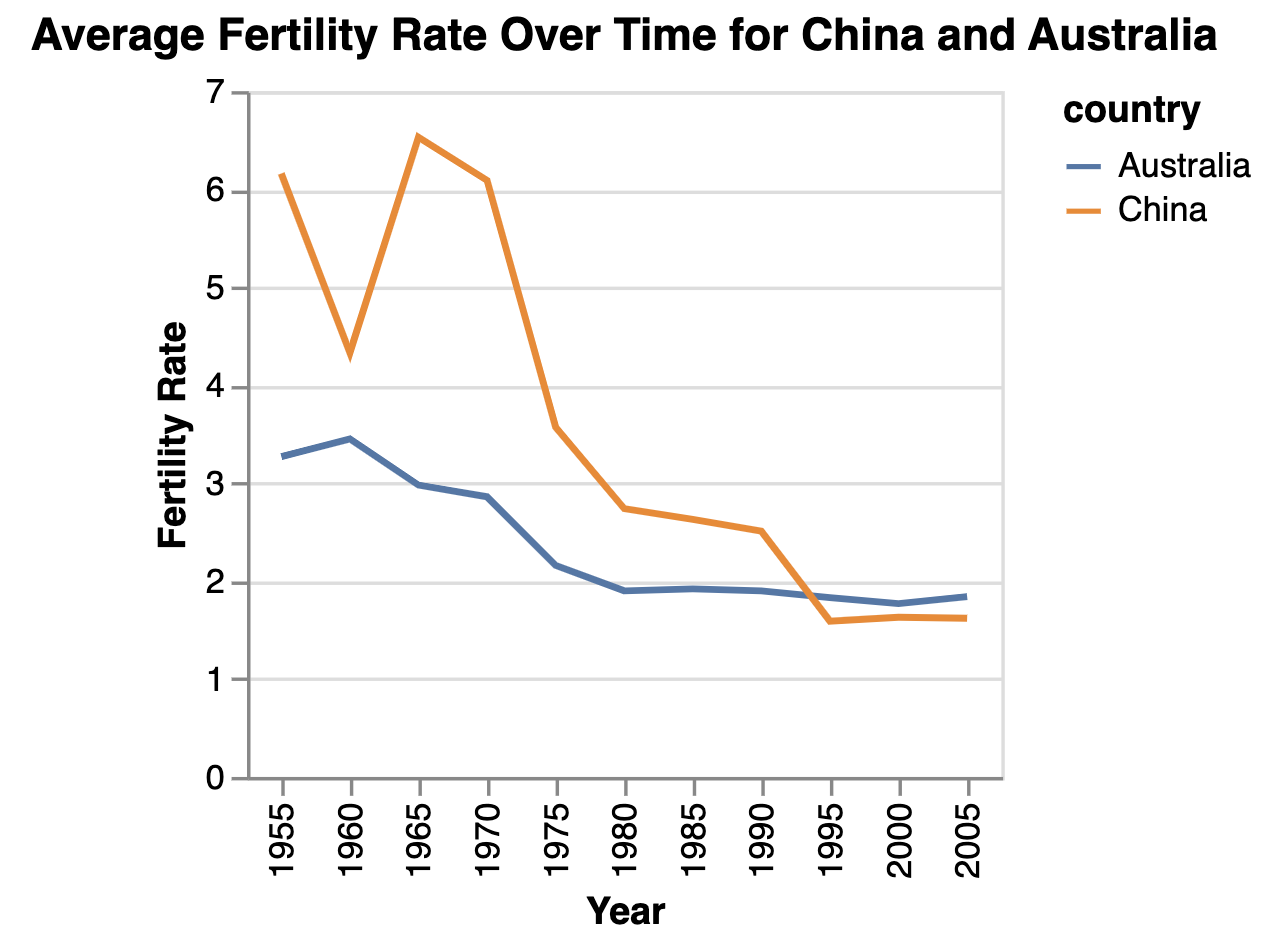}
    \includegraphics[width=0.49 \linewidth]{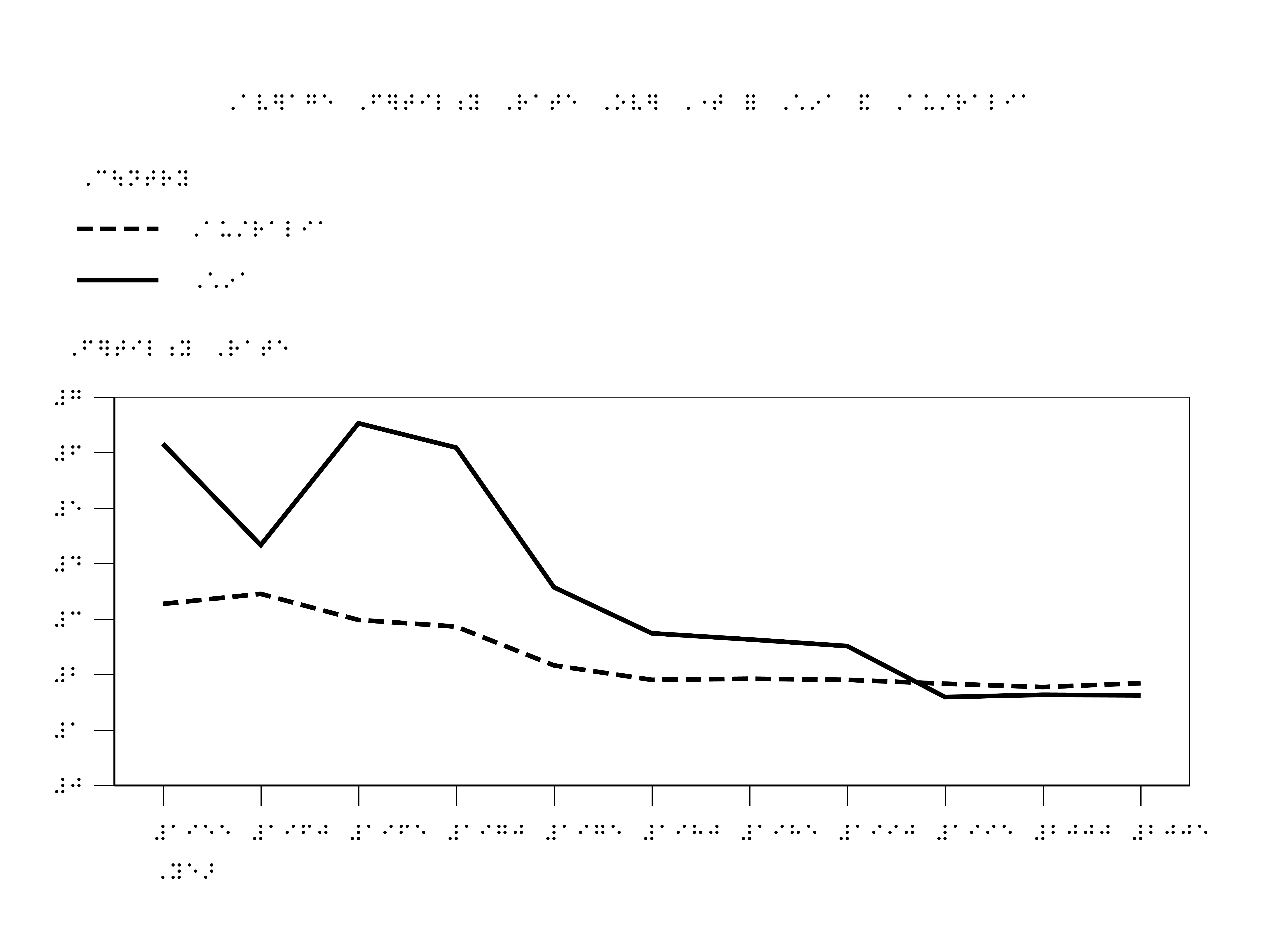}

    \Description{The two images compare a visual and a tactile representation of fertility rate trends for China and Australia over time.
    Left (Visual Graph): A line chart representing the average fertility rate trends for China and Australia from 1955 to 2005. 
    At the top of the chart is the chart title "Average Fertility Rate Over Time for China and Australia," center-aligned. 
    The x-axis is labeled and spans from 1955 to 2005 in increments of 5 years: 1955, 1960, 1965, 1970, 1975, 1980, 1985, 1990, 1995, 2000, and 2005. Axis labels are rotated sideways. Tick marks extend from the x-axis to the x-axis labels. The x-axis is titled "Year" and positioned directly below the x-axis labels.
    The y-axis, positioned on the left, indicates fertility rates ranging from 0 to 7 in increments of 1. Tick marks extend from the y-axis to the y-axis labels. The y-axis is titled "Fertility Rate", rotated sideways and positioned to the left of the y-axis.
    The chart includes a color-coded legend positioned at the top right of the chart area, differentiating the two countries: orange for China and blue for Australia.
    Two lines represent the data for each country: a solid orange line for China and a solid blue line for Australia. The chart shows that China's fertility rate started at a high point near 6 in 1955, sharply declined during the 1960s, rose slightly between 1960 and 1965, and then entered another stage of gradual decline from 1965 onward, stabilizing at around 2 by the 1980s and remaining there through 2005. In contrast, Australia's fertility rate started lower, around 3.5 in 1955, and declined steadily over the decades to just below 2 by 2005.

    Right (Tactile Chart): A chart representing the average fertility rate trends for China and Australia from 1955 to 2005. 
    At the top of the chart is the chart title "Average Fertility Rate Over Time for China and Australia", center aligned.
    The x-axis is labeled in braille and spans from 1955 to 2005 in increments of 5 years. Tick marks extend from the x-axis to the x-axis labels. The x-axis is titled "Year" and positioned right below the x-axis labels. 
    The y-axis, positioned on the left, is also labeled in braille, indicating fertility rates ranging from 0 to 7 in increments of 1. Tick marks extend from the y-axis to the y-axis labels. The y-axis is titled "Fertility Rate" and positioned right above the y-axis. 
    The chart includes a braille legend. The legend is positioned above the chart area, differentiating the two countries by line styles: dashed for China and solid for Australia. 
    Two lines represent the data for each country: a dashed line for Australia and a solid line for China. The chart shows that China's fertility rate started at a high point near 6 in 1955, sharply declined and then started going up from 1960 to 1965, and then entered another stage of gradual decline since 1965. In contrast, Australia's fertility rate started lower, around 3.5 in 1955, and gradually declined to just below two by 2005. }
    \caption{Comparison of visual and tactile charts representing fertility rate trends for China and Australia from 1955 to 2005. This comparison shows design considerations necessary when transforming visual data into tactile formats, such as converting text to braille, adjusting scaling and spacing of chart elements, re-arranging the legend, and substituting visual encodings with tactile encodings.}
    \label{fig:vis_example}
\end{figure*}

Our interviews revealed rich insights on how expert tactile designers conduct their work.
Here, we present an example design workflow synthesized from our interviews and our observation session with the professional tactile graphics designer, using a walkthrough of how a designer would create an example multi-series line chart (\autoref{fig:vis_example}) depicting the average fertility rate of China and Australia from 1955 to 2005.
We synthesized this example workflow by re-watching video recordings of the observational study and systematically analyzing screenshots and intermediate design files produced during the authoring process. This allowed us to identify key decision-making steps, iterative refinements, and recurring patterns, which informed the construction of a generalized workflow representative of tactile chart creation practices.

\subsubsection{Step 1: Familiarizing with complex guidelines.}
Tactile chart designers rely on established guidelines to create accessible and effective graphics. The most authoritative resource in this field is the 2022 Guidelines and Standards for Tactile Graphics published by the Braille Authority of North America (BANA) \cite{brailleauthorityofnorthamericaGuidelinesStandardsTactile2022}. Spanning 426 pages, this comprehensive document provides best practices for everything from basic tactile design to advanced techniques for representing complex information.

Designers usually keep the guidelines open during their work or rely on experienced colleagues for quick consultations. For instance, to design a line graph like the one in our example (\autoref{fig:vis_example}), a designer would look up the relevant section of the guidelines under \emph{Unit 6: Diagrams for Technical Material}, specifically in \emph{6.6 Graphs} \cite{brailleauthorityofnorthamericaGuidelinesStandardsTactile2022}. The designer might begin by reviewing and taking notes on the specific instructions regarding grid spacing, line thickness, and label placement. For example, grid lines should be the least distinct lines in a graph (6.6.4.4); the x-axis (horizontal) and y-axis (vertical) lines must be tactually distinct and stronger than the grid lines, and plotted lines should be the strongest and most tactually distinct lines on the graph (6.6.2.2) \cite{brailleauthorityofnorthamericaGuidelinesStandardsTactile2022}. The designer would take note of these details during the planning stage. 

New designers often find guidelines challenging to navigate due to their highly specific and segmented structure.
For instance, the guidelines are divided into distinct sections for each type of graph---such as line graphs (6.6.4), bar graphs (6.6.6). Information is often repeated across multiple sub-sections, but there are often subtle yet critical differences between graphs that are important to the accessibility of the result. This lack of generalizable principles means designers frequently jump between sections, ensuring they are referencing the correct one for the specific graph type they are working on. 

Furthermore, the guidelines are filled with exceptions and precise details that can be difficult to generalize. For example, bars in tactile bar charts should be between 3/8 inch (1 centimeter) and 1 inch (2.5 centimeters) wide---except for histograms, which should not have spacing between adjacent bars. This specificity makes it harder for designers to develop a broad, transferable understanding of tactile design principles and often requires constant cross-referencing, adding to the already complex process.

\begin{figure*}[h!]
    \includegraphics[width=\linewidth]{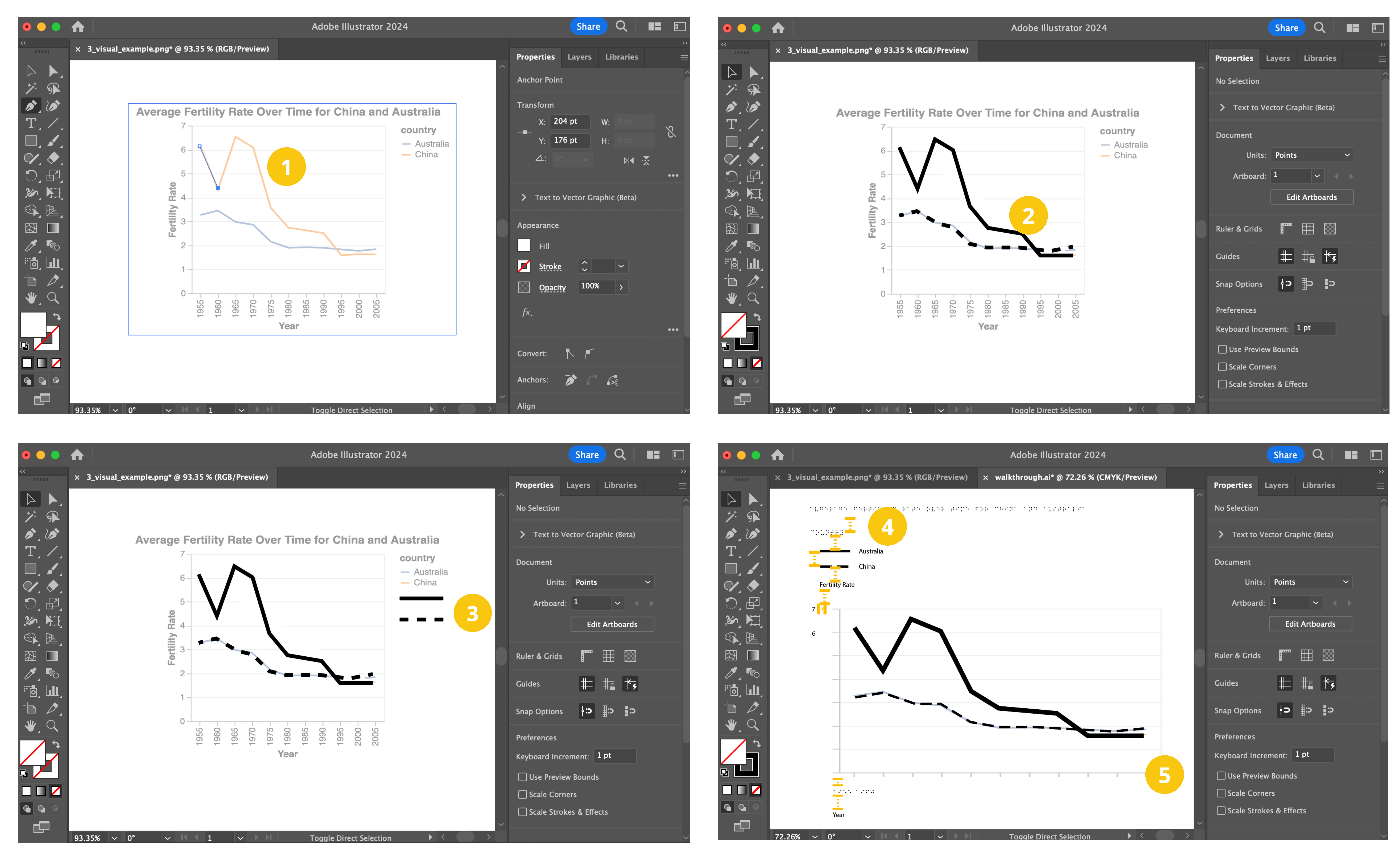}
    \Description{The figure displays a sequence of four screenshots illustrating key steps in the creation of a tactile multi-series line chart using Adobe Illustrator.
    Top left (step 1): A screenshot of Adobe Illustrator displaying the original line chart. The chart is titled "Average Fertility Rate Over Time for China and Australia." The x-axis is labeled "Year," with ticks spanning from 1955 to 2005 at 5-year intervals. The y-axis is labeled "Fertility Rate," with values ranging from 0 to 7 in increments of 1. Two color-coded lines represent data: orange for China and blue for Australia. The legend appears in the top-right corner. Yellow marker 1 is shown on the chart, indicating this is the starting point.  Top right (step 2): The same chart in Adobe Illustrator after the colors of the lines have been converted to black-and-white styles. The orange line (China) is now a solid black line, and the blue line (Australia) is replaced by a dashed black line. This step prepares the chart for tactile conversion by removing color reliance. Yellow marker 2 is placed next to the solid black line, highlighting this transition. Bottom left (step 3): The chart now shows the addition of a legend, added to the right-hand side of the original chart. The solid black line (China) is displayed at the top and the dashed black line (Australia) at the bottom. Yellow marker 3 is placed next to the legends, indicating this modification.  Bottom right (step 4): This screenshot shows the finalized tactile design with braille elements added and spacing adjustments. The x-axis labels (years) and y-axis labels (fertility rates) have been converted into braille text. There are markers illustrating the spacing between the title and the title of the legend, between the legend title and the legend, the legend entries, the y-axis title and the y-axis title, the x-axis and the x-axis label, and the x-axis label and the x-axis title.   Yellow marker 4 is next to the legend area indicating the spacing in the braille legend, while yellow marker 5 is at the end of the x-axis indicating the spacing changes for the x-axis. }
    \caption{Example walkthrough of an expert designer's tactile chart creation process. 1) Tracing a visual reference. 2) Creating tactile encodings. 3) Creating a legend. 4) Adjusting spacing. 5) Scaling the axes to avoid overlaps. }
    \label{fig:3_walkthrough}
\end{figure*}

\subsubsection{Step 2: Constructing the graphic.}

Professional designers often favor vector graphics tools like Adobe Illustrator, valuing their precision and robust capabilities despite a steep learning curve. 
Typically, a designer begins with a visual reference: either using an image, PDF, or PowerPoint slide, or generating a new chart from the data in a tool like Excel. They then import this visual into vector graphics software to trace key elements, including lines, axes, grid lines, and tick marks (\autoref{fig:3_walkthrough}.1).

For our example line chart, the designer differentiates between the two data series by applying distinct line styles for China and Australia (\autoref{fig:3_walkthrough}.2). Since creating textures and line styles from scratch can be time consuming, many professionals utilize Illustrator’s symbols feature to generate reusable graphic assets. Maintaining a customized library of these symbols helps streamline their workflow, though building and managing such libraries can also be labor intensive.

The designer manually selects the grid lines to adjust their properties. They set them to be the least distinct lines relative to the x- and y-axis, which are more tactually distinct. Axes are emphasized to provide a clear reference frame for the reader.
The designer then creates a legend that mirrors the exact line styles and textures used in the chart (\autoref{fig:3_walkthrough}.3). According to guidelines, the legend should be placed before the graph and on the same page when possible.

The designer also adjusts the layout to optimize for the perceptual limitations of touch. Since tactile readers typically use their palms for an overview and fingertips for detail, designers usually scale lines while maintaining the relative spacing ratio between the lines and tactual distinctiveness. In this case, they manually select each grid line in Illustrator and increase the spacing between them. They also resize line styles in the legend to make them more discernible by fingertips. Further adjustments might involve individually selecting and adjusting the spacing between chart elements, such as the chart title and y-axis label, to maintain clarity (\autoref{fig:3_walkthrough}.4).

\begin{figure}[ht]
    \centering
    \begin{subfigure}[t]{0.45\textwidth}
        \centering
        \includegraphics[width=\textwidth]{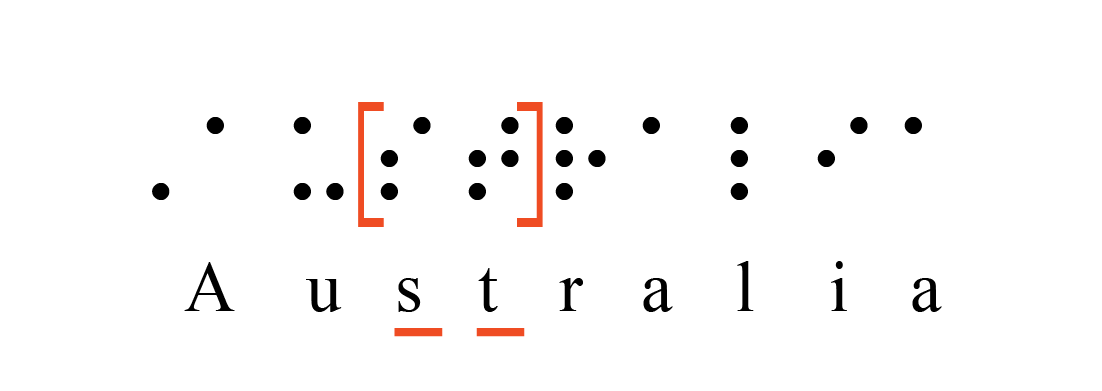}
        \Description{The word Australia in Grade 1 UNICODE braille is rendered using Swell Braille font. The corresponding English letters are shown below the braille letters. Red brackets are around the braille letters s and t. Red underlines are under the letter s and t in English.}
        \caption{Grade 1 uncontracted braille representation of the word ``Australia''. Each letter is an individual cell.}
    \end{subfigure}
    \hfill
    \begin{subfigure}[t]{0.45\textwidth}
        \centering
        \includegraphics[width=\textwidth]{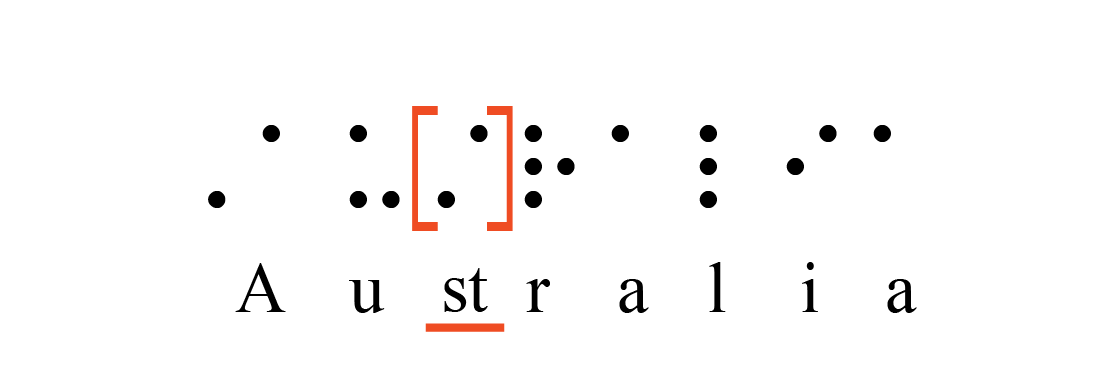}
        \Description{The word `Australia' in Grade 2 UNICODE braille is rendered using Swell Braille font. The corresponding English letters are shown below the braille letters. Red brackets are around the contracted braille "st". Red underline is under the letters "st" in English.}
        \caption{Grade 2 contracted braille representation of the Word ``Australia''. ``st''has been contracted to one cell.}
    \end{subfigure}
    \caption{Braille representations of the word `Australia' in Grade 1 (left) and Grade 2 (right).}
    \label{fig:braille_australia_comparison}
\end{figure}

Designers typically use Duxbury Braille Translation software to convert the chart title, x- and y-axis title and labels, and other text elements into braille. The designer first copies the text and selects the braille code appropriate for the chart audience. For instance, younger children often use Grade 1 uncontracted braille, while adults typically use Grade 2 contracted braille.
For example, the word \texttt{Australia} would be \texttt{,AUSTRALIA} Grade 1 ASCII braille, but would be \texttt{,AU/RALIA} in Grade 2 ASCII braille (the visual difference between these strings rendered in Unicode braille is shown in \autoref{fig:braille_australia_comparison}). In Grade 2 braille, the letters \texttt{st} have been contracted as a single cell. The designer then copies the translated braille back into their graphic design software. Braille introduces spatial constraints because it occupies more space than print text and cannot be resized. The designer then has to re-adjust the layout to fit the braille, including scaling the axes to avoid overlapping braille labels (\autoref{fig:3_walkthrough}.5).

\subsubsection{Step 3: Iterating for context and audience.}
At this design stage, the designer might make further adjustments based on the chart's intended purpose and the reader's familiarity with tactile graphics. For example, the designer might want to simplify the chart by eliminating borders or grid lines that do not contribute to the chart's core message. One educator we interviewed stated that ``less is more,'' and their goal was ``cutting down to the essentials.'' However, the degree of simplification depends on the chart's purpose and audience---designers may retain grid lines if readers are expected to interpret specific data points or add additional vertical grid lines for young readers who need help tracking values across the chart.
Similarly, a designer might adjust line thickness for different readers.
Thinner lines are conducive to precise reading of data values; however, one educator mentioned that precision is less critical for younger students, so they prefer thicker lines because they are easier to locate and trace.
Balancing varying reader needs requires designers to adapt their designs thoughtfully, ensuring each chart is tailored to the user's proficiency and tactile experience while maintaining accessibility and effectiveness.

Once the initial design is complete, the designer needs to iterate on it using specialized preview software to ensure it will effectively translate when physically produced.
For example, designers use Tiger Designer Suite to preview how a design would render on an embosser.
Designers and TVIs often refer to this step as seeing ``how it is going to tiger out''.
To preview the design, the designer first saves their file from Illustrator and then re-opens it in Tiger Designer Suite to preview. While in preview mode, the designer checks the layout to identify potential issues, such as lines being too close or elements overlapping. Depending on the outcome of the preview, the designer then returns to the Illustrator file to make additional edits. They repeat the process above until they are happy with the preview results. Although essential for creating a high-quality tactile chart, previewing and adjusting can be time-consuming and frustrating, which requires a designer to constantly jump back and forth across different software.  

\subsubsection{Step 4: Producing a physical artifact.}

Our interviews primarily discussed two common methods for converting digital chart designs into physical prints.

\textbf{Embossing} involves creating raised lines and textures directly on a sheet of paper, making the final product durable and suitable for long-term use. Embossers like the ViewPlus Delta or Columbia are commonly used for this process. However, embossed lines, especially curved ones, can sometimes appear jagged or segmented, as the embosser cannot always produce perfectly smooth curves. Different graphic embossers can produce varying outcomes due to differences in resolution, dot density, and embossing techniques. Variations in these factors can affect the raised elements' clarity, texture, and precision, leading to discrepancies in how the same image is rendered. Consequently, the tactile quality and readability of the embossed chart can differ depending on the specific embosser used. This is why previewing and adjustments specific to different machines are essential. 

\textbf{Swell forming} uses a special type of paper coated with microcapsules. When black lines are printed on the paper and the sheet is run through a machine that applies heat, the lines swell up, creating raised textures that can be felt. This method allows for smoother curves and more varied textures than embossing, but the resulting tactile graphic may wear down faster with use. Different swell form machines and microcapsule papers can affect the final tactile output because of variations in their sensitivity and embossing capabilities. Swell form machines may differ in temperature and pressure settings, affecting the depth and clarity of the raised features. Similarly, variations in microcapsule paper quality and formulation can influence how well the paper responds to the embossing process, resulting in differences in texture and detail in the final tactile representation.

\subsection{Educator Workflows}

In \autoref{sec:designer-workflows}, we focused mainly on the workflows of our interview group's professional tactile graphic designers and assistive technology specialists. However, we found that our interviewers, who were educators, prioritized different design goals. Where professional designers prioritize precise control and customization, educators instead prioritize affordability, availability, and ease of use.
These priorities result in a different approach to tactile chart creation.
Though we found that professional designers and educators have largely overlapping thought processes, educators typically use more manual approaches to constructing physical charts (Step 2).
This section briefly discusses how an educator's chart construction step differs from a professional designer's.

\begin{figure}[h!]
    \centering
    \begin{subfigure}[t]{0.3\textwidth}
        \includegraphics[width=\textwidth]{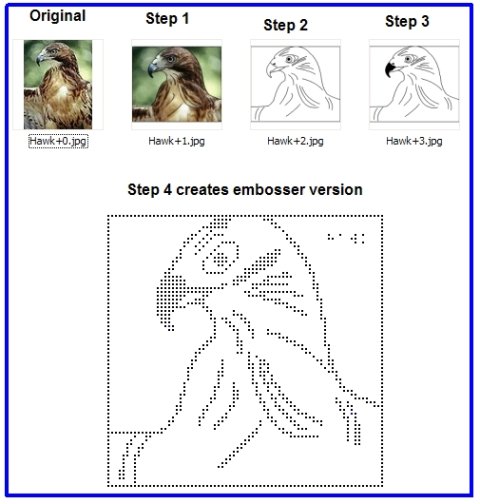}
        \Description{The image illustrates a step-by-step process of converting a visual image of a hawk into a tactile embosser-friendly format, presented in four progressive steps labeled "Original," "Step 1," "Step 2," "Step 3," and "Step 4 creates embosser version." Each step is represented by an image of the hawk at different stages of simplification, arranged in a grid with labels below each step. A detailed breakdown is provided below:         Top Row (Progression Steps 0 to 3):
        1. Original: A full-color, high-resolution image of a hawk. The hawk is facing left with a detailed texture of its feathers and a natural green background. This serves as the starting point for the transformation process.
        2. Step 1: The hawk image has been cropped to focus just on the head of the hawk. Some details in the feathers and edges of the hawk's silhouette are preserved but are less intricate compared to the original image.
        3. Step 2: The hawk image is converted into a black-and-white outline. Only the essential contours of the hawk's head, beak, and feathers are retained. 
        4. Step 3: A clean, line-art version of the hawk is shown, with the difference being the hawk's beak is colored in black.
        Bottom Section (Step 4 creates embosser version): An embossed version of the hawk is displayed, labeled "Step 4 creates embosser version." The hawk is represented as a collection of raised dot patterns suitable for embossing. 
        The dots outline the hawk's head, beak, and feathers, maintaining the key features identified in Step 3. A small braille label is positioned in the top right corner, indicating the hawk's identity in braille text.}
        \caption{Step-by-step process of converting a photo of a hawk into a tactile embosser format using QuickTac \cite{duxburysystemsQuickTacFreeSoftware2024}.}
        \label{fig:3_quicktac}
    \end{subfigure}\hfill
    \begin{subfigure}[t]{0.3\textwidth}
        \includegraphics[width=\textwidth]{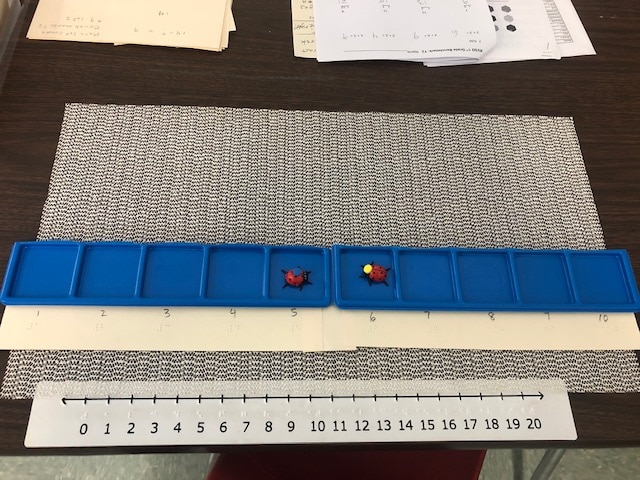}
        \Description{The image shows a classroom setup for a hands-on math activity involving a number line and manipulatives. The setup is on a wooden desk with a textured gray mat underneath. Several items are arranged on the desk:
        Blue Plastic Trays: Two blue rectangular trays are positioned side by side, each divided into five compartments. The compartments are labeled with numbers written in pencil below the trays, starting from 1 on the far left and ending at 20 on the far right (spanning both trays). There are red bug-shaped manipulatives in the fifth and sixth compartments. 
        Number Line: Below the trays is a white number line spanning from 0 to 20, with increments marked at each whole number. The numbers are clearly printed in black.}
        \caption{Hands-on math activity setup featuring blue compartment trays, a number line, and bug-shaped manipulatives \cite{braunerTeachingNumberLine2024}.}
        \label{fig:3_educator_numberline}
    \end{subfigure}\hfill
    \begin{subfigure}[t]{0.3\textwidth}
        \includegraphics[width=\textwidth]{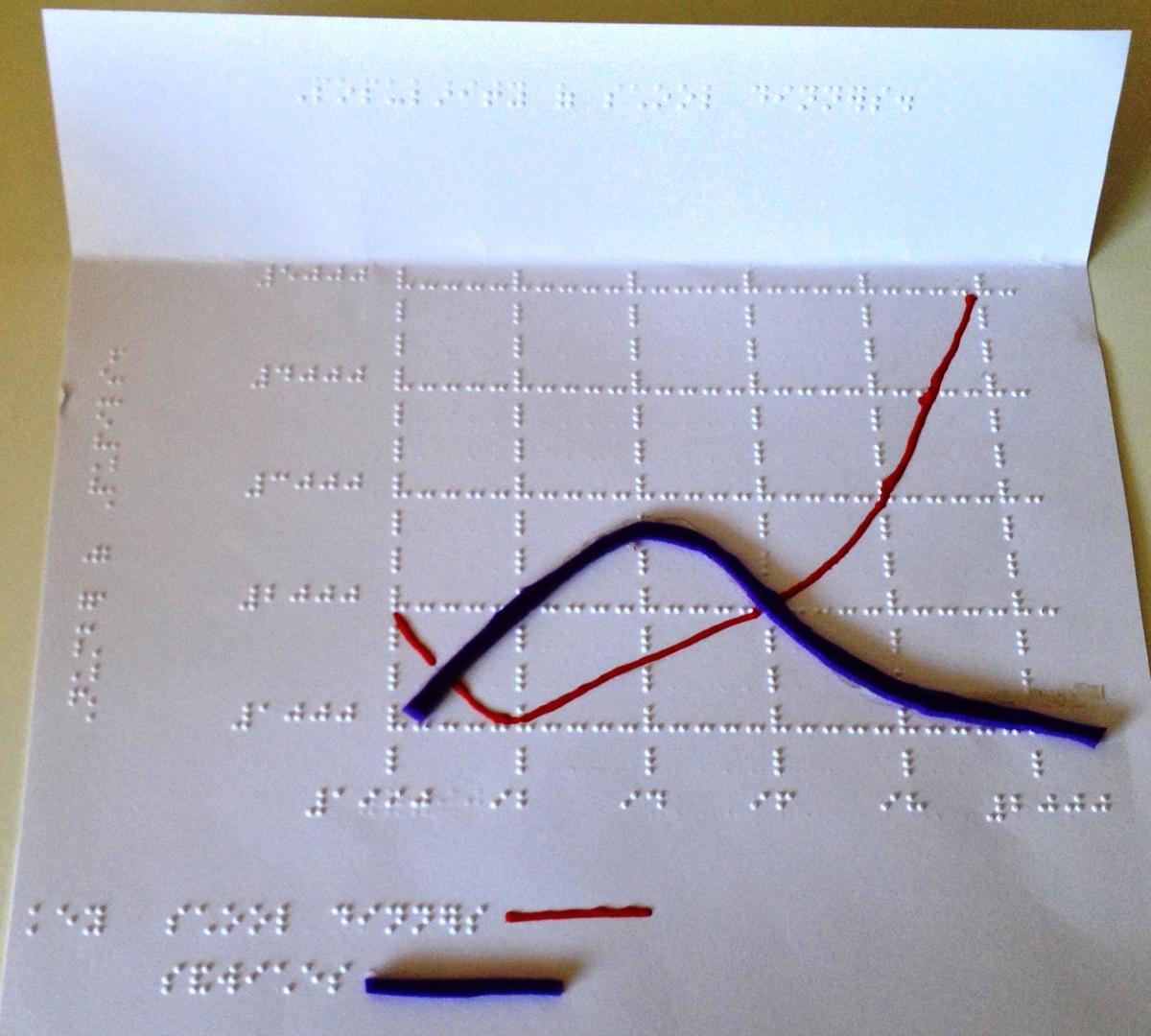}
        \Description{The image shows a tactile graph created on white embossed paper. The graph includes a braille grid, two tactile lines in red and blue, and braille text providing labels and context. 
        Graph Layout: The grid consists of raised braille dots forming horizontal and vertical lines, dividing the graph into evenly spaced intervals. These lines form a rectangular plotting area.
        Braille labels are positioned along the left (y-axis) and bottom (x-axis) edges of the graph. The labels correspond to numeric values, but their specific text is not fully legible in the image.
        Tactile Data Lines:
        Red Line: A thin, raised line beginning near the lower-left corner, gradually curving upward and becoming steeper as it moves to the upper-right corner of the graph. This line represents an increasing trend over time.
        Blue Line: A thicker, more textured tactile line that starts in the middle left of the graph, curves upward to reach a peak and then slopes downward toward the lower-right corner. This line shows a rise and fall pattern.
        Legend: Below the graph, a tactile legend is provided for distinguishing between the red and blue lines. The legend consists of short segments of a red line and a blue line, each aligned with a corresponding braille label that identifies the data they represent. 
        Additional Braille Text: At the top of the page, a title in braille provides a description of the graph. }
        \caption{An example of how a TVI might make tactile charts for the classroom \cite{literacyCreatingLargePrint2014}.}
        \label{fig:3_educator_example}
    \end{subfigure}
    \caption{Educators often use a combination of specialized software, embossing techniques, and hands-on materials to create tactile graphics that convey visual information in accessible formats for blind and low-vision students.}
\end{figure}

In contrast to professional designers, educators tend to prioritize ease of use and speed. One educator with 14 years of teaching experience said that ``time and graphics determine what production is used.'' As a result, they often rely on more straightforward tools like QuickTac (\autoref{fig:3_quicktac}) and physical objects such as pre-printed braille number lines and braille labels (\autoref{fig:3_educator_numberline}). QuickTac \cite{duxburysystemsQuickTacFreeSoftware2024} is free software widely used by educators to trace images and convert them into tactile graphics with basic formatting. It allows users to trace images and import them into the Duxbury Braille Translator (DBT) for braille translation. Although it is tailored for braille graphics, it offers limited customization compared to professional design tools.

Educators we interviewed also shared that they often begin with existing classroom materials, such as worksheets or textbook graphics, as their primary reference. An educator who works at the Marin County Office of Education shared that they ``make tactile charts as often as it comes up in the curriculum.'' Using QuickTac, they trace or draw the graphic and then enhance it with tactile elements created from textured materials, such as braille labels or pre-printed number lines. These elements are manually glued together to form the final tactile chart (\autoref{fig:3_educator_example}). While this approach is fast and practical for classroom use, it is less precise than workflows employed by professional designers, often resulting in lower accuracy and less consistency across charts.

Educators turn to this pragmatic approach because they lack easy-to-use tools that offer both customization and precision and lack time to develop skills in more complex software.
While there might be some pedagogical advantages to using tangible materials like pipe cleaners for line charts, it is likely that many TVIs would like the option to use more automated tools if they were more available and user-friendly. 
The reliance on manual assembly and adaptation of existing materials often reflects a need to bridge the gap between the ideal of professionally designed tactile graphics and the practical constraints of time, resources, and available technology in educational settings.

\subsection{Design Challenges}
Based on our formative interviews with experts, we identified several challenges in designing tactile chart authoring systems.
Here, we describe these design challenges that guided our decision-making as we iterated on the design of Tactile Vega-Lite (TVL).
In \autoref{sec:system-design}, we elaborate on how our design rationale for TVL responds to these challenges.

\textbf{DC1: Guidelines are difficult to learn, apply, and generalize. } 
Existing tactile graphics guidelines are highly prescriptive in that they often consist of lengthy and detailed lists of criteria that should be met in a tactile chart.
While this level of detail makes it clear how designers should implement guidance and makes it easy to tell when charts are not complying with guidelines, there are also drawbacks to this approach.
For instance, designers can sometimes find these guidelines challenging to learn independently, particularly when limited examples or experienced mentors are available. Even when designers are familiar with guidelines, remembering and applying them consistently can be difficult.
Additionally, correctly implementing the guidelines in a design can be extremely tedious and error-prone, involving numerous manual adjustments to account for fine details. 

Further, the guidelines are non-exhaustive; they only cover a small subset of data visualizations. For example, guidelines provide detailed instructions on creating bar charts, line charts, pie charts, and scatter plots but do not cover strip plots, bubble plots, grouped bar charts, and area charts, all of which are commonly used in practice by journalists and other visualization designers.
Generalizing guidelines to other chart forms is difficult because they focus on prescriptive guidance rather than higher-level design principles.
When the guidelines do not cover a certain chart type or situation, designers must make judgment calls that typically require years of experience or extensive practice.

\textbf{DC2: Existing tools require high skill even to create basic charts.} Designers face a trade-off between complexity and control. 
Tools like Adobe Illustrator that provide extensive customization options are often challenging to master, especially for those without specialized training. 
While there are ways to streamline workflows—such as using shortcuts or creating reusable symbols and textures—newcomers are often unaware of these tips and tricks. 
Mastering these techniques requires time and experience, which many novice designers or those new to tactile graphics lack.
On the other hand, tools such as SVGPlott \cite{engelSVGPlottAccessibleTool2019} that automate many aspects of the process offer limited customization, restricting designers' ability to create tailored tactile graphics that meet specific user needs. 
The lack of intermediate tools—those that balance ease of use with meaningful customization—forces designers to choose between two extremes: highly complex, professional-grade software or overly simplified, automated solutions. This trade-off can be particularly frustrating when trying to meet the needs of diverse audiences, from beginners learning to read tactile graphics to advanced users requiring highly detailed and accurate representations. 

\textbf{DC3: Tools lack affordances specific to tactile chart design.} 
Most professional designers use general-purpose vector graphics software, like Adobe Illustrator, to create tactile charts.
Because these tools are not designed specifically for tactile chart design, workflows can become unwieldy as designers translate tactile chart concepts into low-level graphical primitives.

For example, when a designer wants to make a change to a data-driven tactile mark, they will typically want to apply this change to every mark.
However, they are forced to manually select and adjust each element, as most tools lack the ability to propagate changes across identical elements easily. This manual effort adds considerable time to the workflow.

Moreover, designers often transfer data between different platforms; for instance, they create a chart in Excel and then move it to a vector design tool like Illustrator to trace it. 
Designers must also preview the tactile chart in separate software to ensure it renders correctly for the intended production method (e.g., embossing or swell-forming). If the preview reveals issues, they must return to Illustrator, make edits, and repeat the process, leading to a frustrating cycle of back-and-forth adjustments.
This cross-platform transfer introduces the risk of losing accuracy in the tactile representation and is only necessary because of the lack of domain-specific tool support.

These inefficiencies underscore the need for tools designed for tactile chart creation, with built-in affordances for managing tactile-specific assets, propagating changes across elements, and streamlining the iterative process.

\section{System Design}
\label{sec:system-design}

Tactile Vega-Lite (TVL) is a set of extensions to the Vega-Lite grammar that enables rapid prototyping of tactile charts.
We chose to extend Vega-Lite because we wanted to express a wide variety of tactile charts using a concise set of primitives, which is made possible by Vega-Lite's Grammar of Graphics (GoG)-based approach.
We also hoped to enable visualization authors to easily specify default charts that are informed by best practices by making small edits to existing visual specifications without requiring tactile expertise.
Vega-Lite's design and use of defaults lends itself to this approach.

To design TVL, we alternated between prototyping the desired syntax of TVL as fragments of JSON and prototyping a TVL compiler that could produce SVG tactile charts from our syntax.
Our prototype TVL compiler is open source, and its code can be found at \url{https://github.com/mitvis/tactile-vega-lite}.

We also conducted informal testing of TVL charts by producing them on swell paper and testing with blind colleagues---one proficient in tactile graphics and one less familiar---to ensure that they could successfully interpret chart contents and answer data-related questions. They also provided feedback on design choices and defaults, which then in turn informed our language design.

In this section, we first introduce TVL's tactile abstractions and default specifications. Then, we discuss our design rationale and how it addresses the design challenges we identified.

\subsection{Tactile Vega-Lite Abstractions}

TVL extends Vega-Lite with new abstractions that address design considerations specific to tactile charts.
These abstractions enable the specification of tactile encodings, braille usage, navigational aids, and layout configuration.
For each specification, TVL provides default values (\autoref{sec:default-values}) motivated by best practices in tactile design.
In this section, we review each language extension, addressing how they build on Vega-Lite's existing grammar of graphics and explaining our choice of defaults.

\begin{figure*}
    \includegraphics[width=\textwidth]{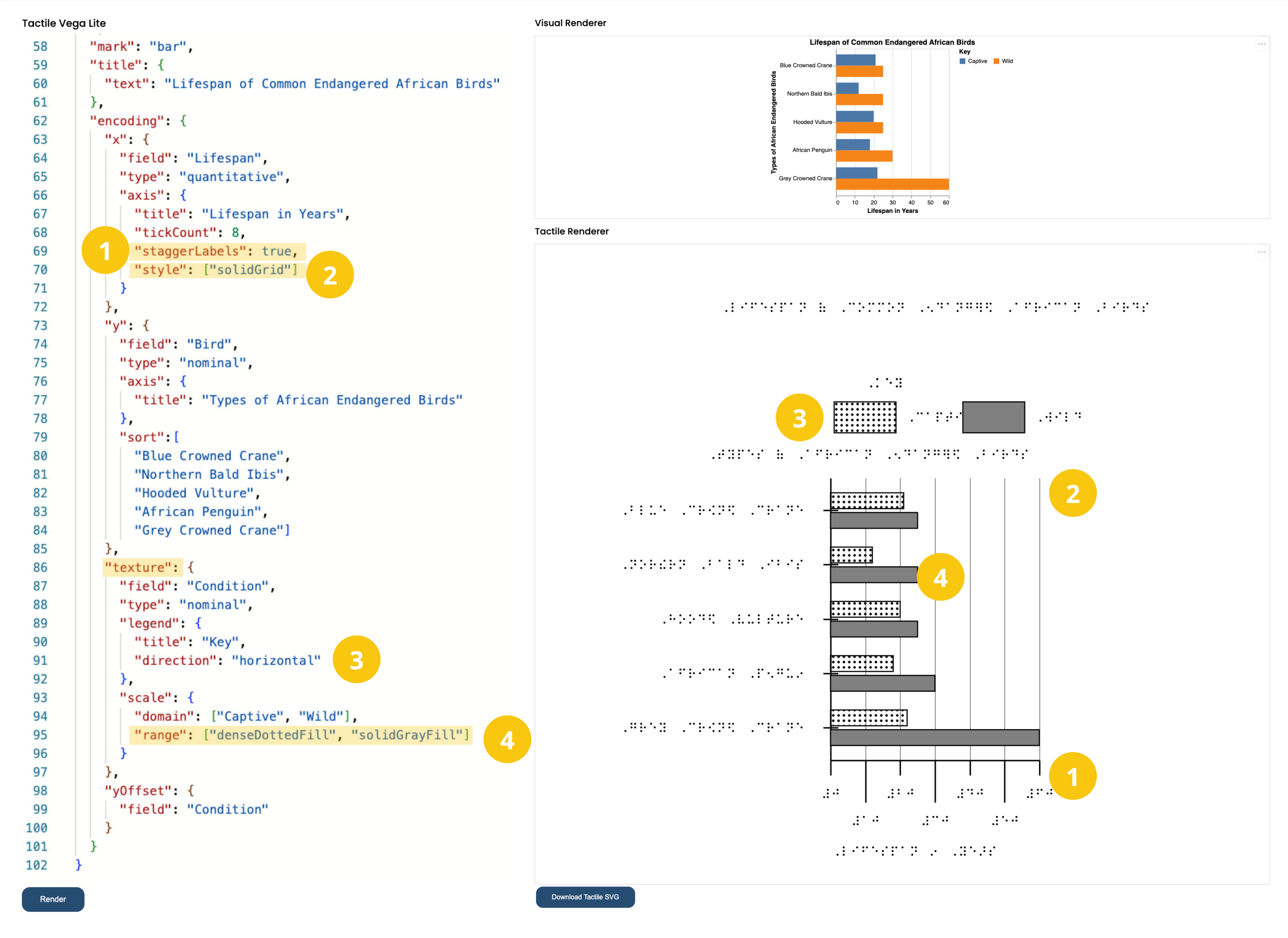}
    \Description{The interface has three primary sections: a code editor on the left, a visual chart renderer on the top right, and a tactile chart renderer on the bottom right. Yellow numbered markers (1 through 4) are used to highlight specific elements in both the code and the rendered charts. Below is a detailed breakdown of each section:    
    Left: Tactile Vega-Lite Code Editor
    The left side shows a JSON-based configuration file written in the Tactile Vega-Lite format, used to define both visual and tactile bar charts. The JSON file includes the following key elements:
    1. Chart Title: "Lifespan of Common Endangered African Birds".
    2. X-axis ("x" Field): Maps the "Lifespan" field to the horizontal axis, labeled "Lifespan in Years". Includes properties such as "tickCount": 8, "staggerLabels": true (this line of code is highlighted in yellow), and "style": "solidGrid" (this line of code is highlighted in yellow).
    3. Y-axis ("y" Field): Maps the "Bird" field to the vertical axis, labeled "Types of African Endangered Birds". A "sort" field specifies the order of bird names: "Blue Crowned Crane", "Northern Bald Ibis", "Hooded Vulture", "African Penguin", "Grey Crowned Crane".
    4. Texture Encoding ("texture" Field, highlighted in yellow in the code editor): Encodes the "Condition" field to apply textures for "Captive" and "Wild" conditions. Includes a "legend" with a horizontal layout, assigning textures like "denseDottedFill" for "Captive" and "solidGrayFill" for "Wild" (this line of code is highlighted in yellow).
        Top Right: Visual Chart Renderer
    The top-right section displays a visual bar chart based on the JSON configuration:
    - The chart is titled "Lifespan of Common Endangered African Birds".
    - The x-axis is labeled "Lifespan in Years" and spans from 0 to 60 years.
    - The y-axis is labeled "Types of African Endangered Birds" and lists bird names such as "Blue Crowned Crane" and "African Penguin". It is positioned vertically to the left of the axis.
    - Bars are horizontally grouped and color-coded: Orange represents "Wild". Blue represents "Captive".
    - The legend, located at the top right of the chart, identifies the colors. The legend entries are on the same row. A blue square is followed by "Captive," and then an orange square is followed by "Wild." 
      Bottom Right: Tactile Chart Renderer
    The bottom-right section displays a tactile bar chart rendered for visually impaired users. Key elements include:
    1. Embossed Braille Labels: The chart features braille text along both axes. The x-axis (horizontal) is labeled "Lifespan in Years" with tick marks from 0 to 60. The y-axis (vertical) lists the bird names in braille, such as "Blue Crowned Crane" and "Hooded Vulture".
   2. Tactile Bar Representation: Each bar is filled with a unique texture:
     - A dotted texture represents "Captive".
     - A solid texture represents "Wild".
     3. Gridlines: Solid tactile gridlines are perpendicular to the x-axis.
   4. Tactile Legend: Positioned above of the tactile chart. It includes a dotted bar labeled "Captive" and a solid bar labeled "Wild", both annotated with braille.
     Highlighted Markers (1 to 4):
   1. Marker 1: Highlights the "staggerLabels": true property in the code, which staggers every other x-axis label such that some labels are on one line and others are on another line. The corresponding marker 1 on the tactile chart is next to the staggered x-axis labels. 
   2. Marker 2: Highlights the "solidGrid" grid lines, which are vertical lines going through the chart. The corresponding marker 2 on the tactile chart is positioned next to the rightmost grid line near the top of the chart.  
   3. Marker 3: Highlights the direction of the legend "horizontal", indicating the legend entries horizontally displayed. The corresponding marker 3 is positioned to the left of the first legend entry above the tactile chart. 
   4. Marker 4: Refers to the "texture" field in the code, which maps the "Condition" data to tactile textures for "Captive" and "Wild". The corresponding marker 4 is positioned next to the solid bar next to the hooded vultures. } 
   \caption{Tactile chart creators can use the TVL code editor to customize key properties such as axis labels, tick marks, sorting, and texture encodings. The rendered outputs on the right showcase the parallel visual and tactile representations. Highlighted code snippets in yellow show properties that were added in TVL.}
\end{figure*}

\subsubsection{Tactile Encodings}

\begin{figure*}
    \centering
    \includegraphics[width=0.98\textwidth]{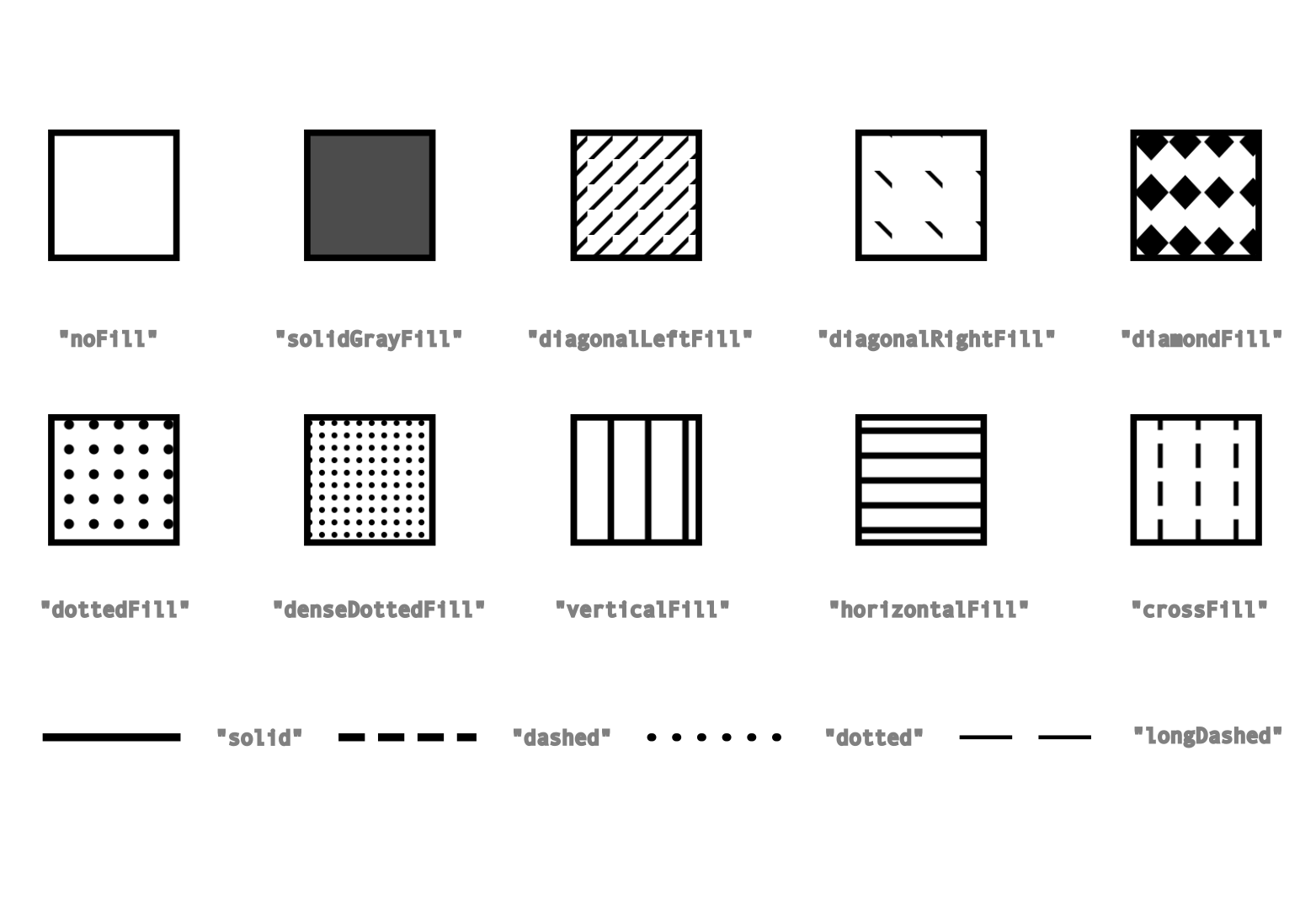}
    \Description{The image displays a set of textures and line styles commonly used in design or data visualization for accessibility. Each texture is shown in a square. There are 5 columns of squares and 2 rows for a total of 10 textures. The name of each texture fill is displayed below each square. 
      The description moves left to right, top to bottom:
    1. "noFill": A square with no fill; only a black outline is visible.
    2. "solidGrayFill": A square with a uniform gray fill and a black outline.
    3. "diagonalLeftFill": A square filled with diagonal lines slanting from top-right to bottom-left, with a black outline.
    4. "diagonalRightFill": A square filled with diagonal lines slanting from top-left to bottom-right, with a black outline.
    5. "diamondFill": A square filled with a repeating diamond pattern in black on a white background, with a black outline.
    6. "dottedFill": A square filled with evenly spaced black dots on a white background with a black outline.
    7. "denseDottedFill": A square with a denser arrangement of black dots compared to "dottedFill," on a white background, with a black outline.
    8. "verticalFill": A square filled with vertical black stripes spaced evenly apart, with a black outline.
    9. "horizontalFill": A square filled with horizontal black stripes spaced evenly apart, with a black outline.
    10. "crossFill": A square filled with dashed vertical black lines evenly spaced on a white background, with a black outline.
    Below the textures are line styles. There are four line styles displayed in a row in total. The names of each line style follow the line style. 
    Line styles are as follows:
    1. "solid": A horizontal solid black line with no breaks.
    2. "dashed": A horizontal line made of short black dashes with equal spacing in between.
    3. "dotted": A horizontal line made of evenly spaced black dots.
    4. "longDashed": A horizontal line made of longer black dashes, longer than the "dashed" line, spaced further apart.}
    \caption{A collection of textures and line styles designed for accessibility in charts and diagrams. The textures include various fill patterns such as solid gray, diagonal lines, dots, and grid patterns, while the line styles include solid, dashed, dotted, and long-dashed lines.}
    \label{fig:texture_sampler}
\end{figure*}

In visualization, encodings are mappings of data fields to perceptual properties such as color, size, and position.
Tactile Vega-Lite shares many encodings with base Vega-Lite, such as the \texttt{x} and \texttt{y} positional encodings, and encodings like \texttt{size} and \texttt{shape}.
However, it does not include encodings for visual-only properties like \texttt{color} or \texttt{opacity}.
Instead, TVL introduces a \texttt{texture} encoding that maps a data field to a set of patterns that are perceptually distinct by touch.
Drawing on empirical research and guidelines on pattern sets \cite{edmanTactileGraphics1992, prescherConsistencyTactilePattern2017, brailleauthorityofnorthamericaGuidelinesStandardsTactile2022, watanabeEffectivenessTactileScatter2018, watanabeTexturesSuitableTactile2018}, we designed a default texture palette for TVL with 10 fill textures (\autoref{fig:texture_sampler}). 
 
By default, the system automatically assigns discrete texture encodings to a subset of textures from the built-in palette. If only one data value is present in the encoding, we assign \verb|solidGrayFill| as the default texture because empty bars can complicate the tactile differentiation between bars and white space around them \cite{engelAnalysisTactileChart2017, challisDesignPrinciplesTactile2001}. 
When more than five textures are used in the chart, the system alerts the author and recommends that they consider alternate encodings. This is because readers may have difficulty learning and recalling the mapping between textures and data values when more than five textures are used.

A designer can override the default texture by specifying an alternate set of named textures from our set of ten (\autoref{fig:example-texture}).
Though Vega-Lite can technically support arbitrary SVG textures, they are difficult to specify concisely --- for instance, a variety of colors can be expressed concisely through RGB notation, but there is no analogous notation for texture.
Thus, we leave specialized language support in TVL for arbitrary textures for future work.

Where textures define a tactile pattern for an area, TVL uses the existing \texttt{strokeDash} encoding to define tactile patterns for lines and \texttt{strokeWidth} encoding to control the thickness of the line. 
TVL provides a set of pre-defined \texttt{lineStyles}, which include \texttt{dashed}, \texttt{solid}, \texttt{dotted}, \texttt{longDashed} (\autoref{fig:texture_sampler}).  
These line styles, combined with variations in \texttt{strokeWidth}, enable designers to create a wider range of tactile line patterns.
Like with texture, the system prompts the author to consider creating multiple charts when more than four different line styles are present. 

By default, when there is only a single plotted line, the system always chooses the solid line style because it is the most easy-to-recognize tactile pattern, minimizing confusion for readers unfamiliar with other line styles. In a multi-series line chart, the system will always choose the solid line as one of the line styles and pair other styles with it because the contrast between a solid line and other patterns, such as dashed or dotted lines, is higher. Designers can customize arbitrary line styles by specifying the \verb|strokeDash| encoding using SVG's standard dash array notation.

\subsubsection{Braille Usage}
Braille uses six raised dots arranged in a systematic pattern of two columns with three dots each, forming what is known as a Braille cell \cite{internationalcouncilonenglishbrailleicebUnifiedEnglishBraille2024}. Unified English Braille (UEB) is the standardized system of English Braille \cite{internationalcouncilonenglishbrailleicebUnifiedEnglishBraille2024}, which includes uncontracted braille (Grade 1) and contracted Braille (Grade 2). Uncontracted braille is typically easier for beginners to learn but requires more space, limiting its use primarily to early elementary education. Contracted braille, the more advanced and widely used form, incorporates abbreviations, contractions, and shorthand symbols to represent common words, parts of words, or groups of letters. The numeric indicator in braille is a specialized symbol that signals the subsequent braille cells should be interpreted as numbers rather than letters. This indicator is used only once in sequences with multiple digits before the first digit. This distinction is crucial in contexts like tactile charts, where accurate representation of numbers is frequently required. 

TVL implements braille translation using the open-source library LibLouis \cite{liblouisLiblouisOpensourceBraille2024}, which is well-regarded within the braille transcriber community. The default settings in TVL use versatile and high-quality \verb|Swell Braille| font at \verb|24pt| a font size for readable Swell Braille in tactile graphics, applying Grade 2 braille with the \verb|en-ueb-g2.ctb| translation table from LibLouis. Given that braille fonts can vary depending on production methods, designers can choose from additional braille fonts, such as \verb|California Braille|, which refers to a specific braille dot and cell spacing standard unique to California, designed for compliance with state regulations for public signage, or \verb|Braille29|, which is used to create tactile graphics embossed using the Tiger embosser, available in many languages. 

\subsubsection{Navigational Aids}

Vega-Lite provides navigational aids such as grid lines, axes, and axis ticks, which help users interpret data by offering reference points for scale and structure. These visual aids assist with understanding relationships between data points and orienting users within a chart.
Tactile Vega-Lite builds upon these elements by introducing a clear information hierarchy among navigational aids, which assist users in orienting and navigating the chart's layout. This hierarchy organizes the relative prominence of grid lines, axes, tick marks, and data lines, ensuring that critical information is easy to detect while supporting elements remain unobtrusive. 

For tactile charts, clear navigational aids are critical due to differences in visual and tactile wayfinding and exploration. According to Lucia Hasty, co-author of the BANA Guidelines and Standards for Tactile Graphics, unlike sighted individuals who typically grasp visual graphics in a ``whole-to-part'' manner, blind readers usually explore tactile graphics in a ``part-to-whole'' sequence \cite{hospitalTipsReadingTactile2024}. They begin by examining individual elements of the graphic and gradually piece them together to understand the overall structure and context. Tactile readers rely heavily on anchors---specific points on a graphic that they can consistently refer back to while navigating the rest of the image.
These elements facilitate the reading and comparison of values and serve as navigational breadcrumbs, allowing readers to trace their path back to previous points. It is crucial to balance these enhancements with minimizing clutter and maintaining clarity \cite{engelImproveAccessibilityTactile2017}. 

Information hierarchy for navigational aids is expressed through variations in line styles, thickness, and placement \cite{brailleauthorityofnorthamericaGuidelinesStandardsTactile2022}. According to guidelines, grid lines should be the least distinct lines in a chart. The x-axis and y-axis should be tactually distinct and stronger than the grid lines. Tick marks may have the same line strength as the axes or have a line strength stronger than grid lines and weaker than axes. 

However, research has not reached definitive conclusions on how to judge a line as more or less tactually distinct.
For example, it is unclear whether line style or thickness plays a more significant role in tactile distinctiveness (e.g., whether a thinner solid line is more or less distinct than a thicker dotted line).
TVL offers various customization options for designers to adjust on a case-by-case basis. 

TVL uses default configurations to express the navigational aid hierarchy by varying only the line thickness. 
Conventionally, solid lines represent primary paths, while dashed or dotted lines indicate secondary or auxiliary lines. Line thickness can also convey hierarchy, with thicker lines representing more prominent features or paths and thinner lines for less significant ones. Additionally, the placement of these elements in relation to other tactile encodings, such as texture or braille labels, helps users orient themselves within the information structure. 
TVL provides customizable ways to express this hierarchy, allowing designers to adjust line thickness, style, and placement to meet specific tactile needs. 

\paragraph{Grid lines.} Grid lines provide a reference framework that helps readers track the positions of data points relative to the axes. Grid lines are useful when readers need to track values accurately, such as in bar or line charts. Research shows grid lines can improve readability, reading speed, and accuracy, though they can increase reading time \cite{ledermanTangibleGraphsBlind1982, aldrichTangibleLineGraphs1987, barthTactileGraphicsGuidebook1982, goncuUsabilityAccessibleBar2010, goncuTactileChartGeneration2008}. TVL applies grid lines by default to quantitative axes based on the chart's encoding. A designer can customize grid line styles and change the background/foreground stacking order (\autoref{fig:example-gridlines}).

\paragraph{Axes and axis ticks.} The x- and y-axis provide a frame of reference for tactile wayfinding by defining the bounds of the chart. Guidelines suggest that they should be tactually distinct and stronger than the grid lines but less than plotted lines \cite{brailleauthorityofnorthamericaGuidelinesStandardsTactile2022}. 
For tactile readers, axis ticks serve a dual function: on quantitative scales, they facilitate the reading of specific values and maintain orientation, while on categorical axes, they enable the identification of mark positions and quantity through tactile scanning \cite{engelAnalysisTactileChart2017}. They also aid readers in differentiating chart elements, discerning units, and accurately associating labels with the corresponding axes. 

TVL's \verb|staggerLabels| parameter staggers the x-axis labels, placing alternating values one or two lines below the x-axis --- as suggested by the BANA guidelines \cite{brailleauthorityofnorthamericaGuidelinesStandardsTactile2022}. When \texttt{staggerLabels} is set to \texttt{true}, a lead line extends the tick marks to the staggered labels on the lower level (\autoref{fig:example-stagger}). \textit{Lead lines} are connecting lines that link data points to axes or other reference points and labels, making it easier for tactile readers to follow relationships between elements. By default, \texttt{staggerLabels} is set to \texttt{"auto"}; in this case, the system staggers axis labels when the length of the labels exceeds the threshold.

\subsubsection{Layout Configuration}

Layout refers to chart elements' relative positioning, spacing, alignment, and orientation. Tactile readers often employ a systematic scanning technique,  using two hands to gather information from the top to the bottom of a page. This scanning method helps identify key features such as titles, labels, and other critical elements before diving into the finer details. A good layout uses alignment and negative space to enhance readability and comprehension while respecting readers' reading habits. 
Here, we discuss TVL's use of positioning, spacing, and alignment to manage layout considerations.

\begin{figure*}[!ht]
    \centering
    \begin{subfigure}[t]{0.44\textwidth}
        \centering
        \includegraphics[width=\textwidth]{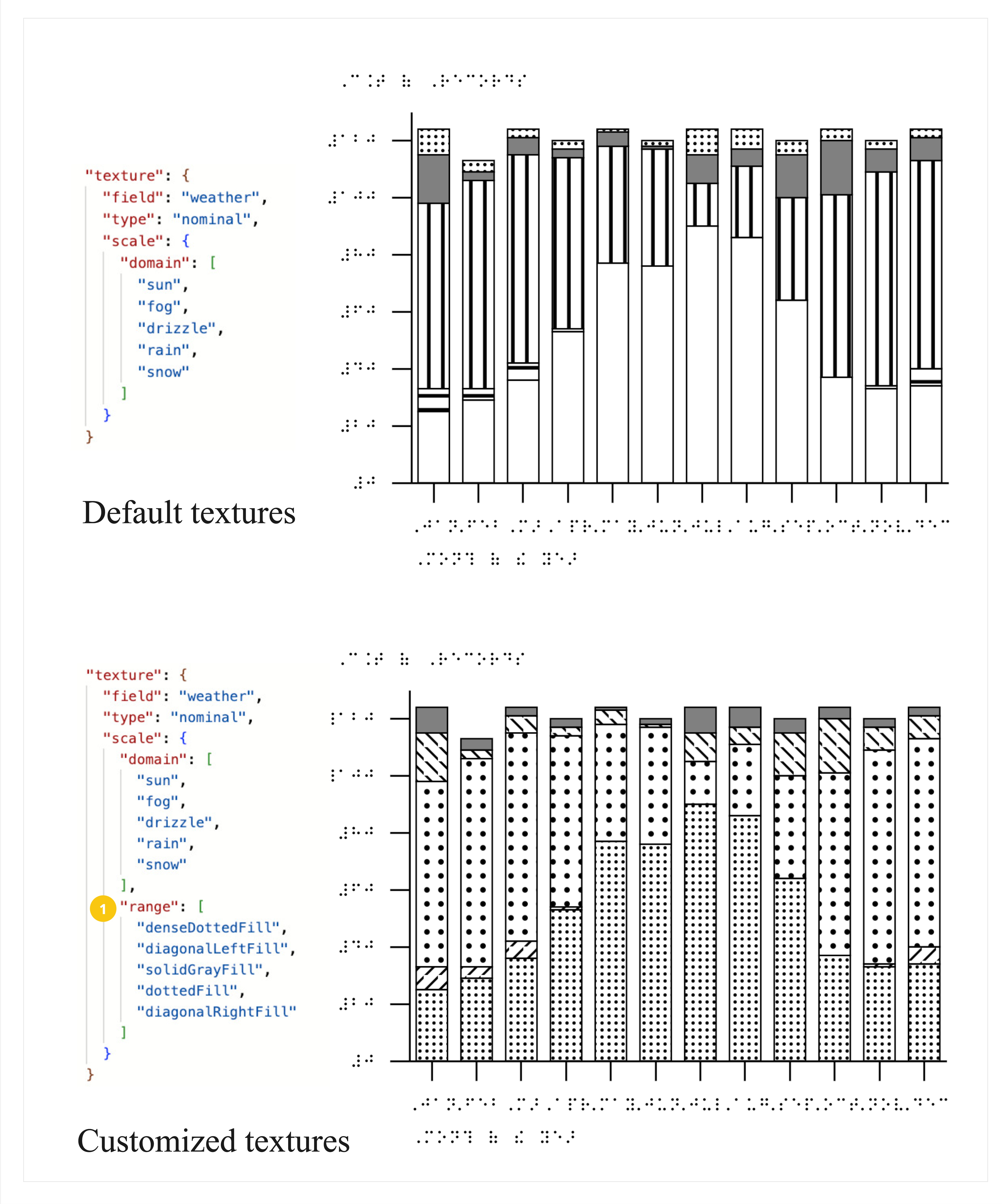}
        \Description{The image compares two tactile bar charts illustrating the use of default and customized textures to encode categorical data. The image consists of two main sections: the top chart uses default textures, and the bottom chart uses customized textures. Each section includes a JSON code snippet on the left that defines the texture properties and a tactile bar chart on the right.
        Top Section: Default Textures
        - Code Snippet: The JSON code snippet defines the "texture" property with a "field" set to `"weather"`, a "type" of `"nominal"`, and a "scale.domain" that maps five categories (`"sun"`, `"fog"`, `"drizzle"`, `"rain"`, `"snow"`) to default textures.
        - Tactile Chart: A stacked bar chart. The textures in each stacked bar chart from bottom to top are empty fill, horizontal fill, vertical fill, solid gray fill, and densely dotted fill. 
        Bottom Section: Customized Textures
        - Code Snippet: The JSON code snippet adds a `"range"` property to customize the textures for the same weather categories. The textures in each stacked bar chart from bottom to top are densely dotted fill, diagonal right fill, sparsely dotted fill, diagonal right fill, and solid gray fill. 
        Braille labels annotate the axes similarly to the top chart.
        Yellow marker 1 is displayed on the bottom section next to the "range" snippet of code. 
        }
        \caption{Comparison of default and customized textures in tactile bar charts.}
        \label{fig:example-texture}
    \end{subfigure}
    \hfill
    \begin{subfigure}[t]{0.44\textwidth}
        \centering
        \includegraphics[width=\textwidth]{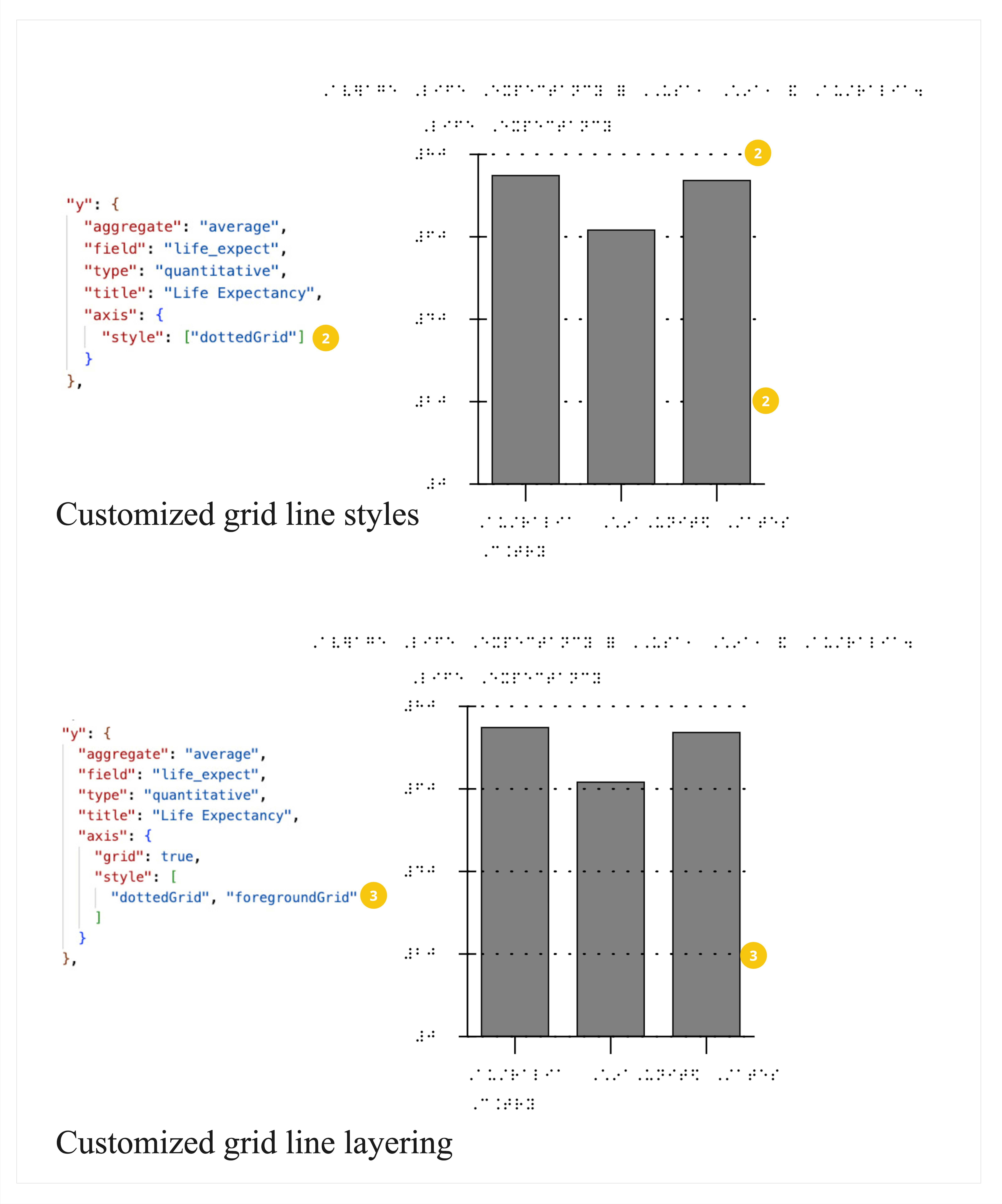}
        \Description{The image compares two examples of customized grid line styles and layering in tactile bar charts demonstrated using a Tactile Vega-Lite configuration. The image is split into two sections, each containing a snippet of JSON code on the left and the corresponding tactile-rendered bar chart on the right.
                Top Section: Customized Grid Line Styles
        - JSON Code: The JSON snippet defines the y-axis field `"life_expect"` with an aggregation of `"average"`. The axis style is set to `"dottedGrid"`, indicated by the `"style": ["dottedGrid"]` property.
        - Tactile Chart: The chart shows three vertical bars representing life expectancy averages, with dotted horizontal grid lines spanning the chart. Where the grid line and bar intersect, the grid line is not visible. Braille labels annotate the axes, providing quantitative values on the y-axis and categorical labels on the x-axis. Yellow marker 2 highlights the use of dotted grid lines in both the JSON snippet and the tactile chart.
        Bottom Section: Customized Grid Line Layering
        - JSON Code: The JSON snippet adds an additional style layer to the y-axis grid. The `"grid": true` property is enabled, and the `"style"` property includes both `"dottedGrid"` and `"foregroundGrid"` for layered grid customization.
        - Tactile Chart: The chart displays three vertical bars similar to the top example, but now where the bar intersects the grid line, the grid line is visible. Yellow marker 3 highlights the JSON configuration and the tactile rendering of the layered grid.
Yellow marker 2 is shown in the top chart, next to the "dottedGrid" code snippet, next to the top grid line, and next to the rightmost bar. 
        Yellow marker 3 is shown in the bottom chart, one next to the style code snippet, one next to the right most bar on the last gridline.  
        }
        \caption{Example of gridlines passing through the chart.}
        \label{fig:example-gridlines}
    \end{subfigure}
    \begin{subfigure}[t]{0.44\textwidth}
        \centering
        \includegraphics[width=\textwidth]{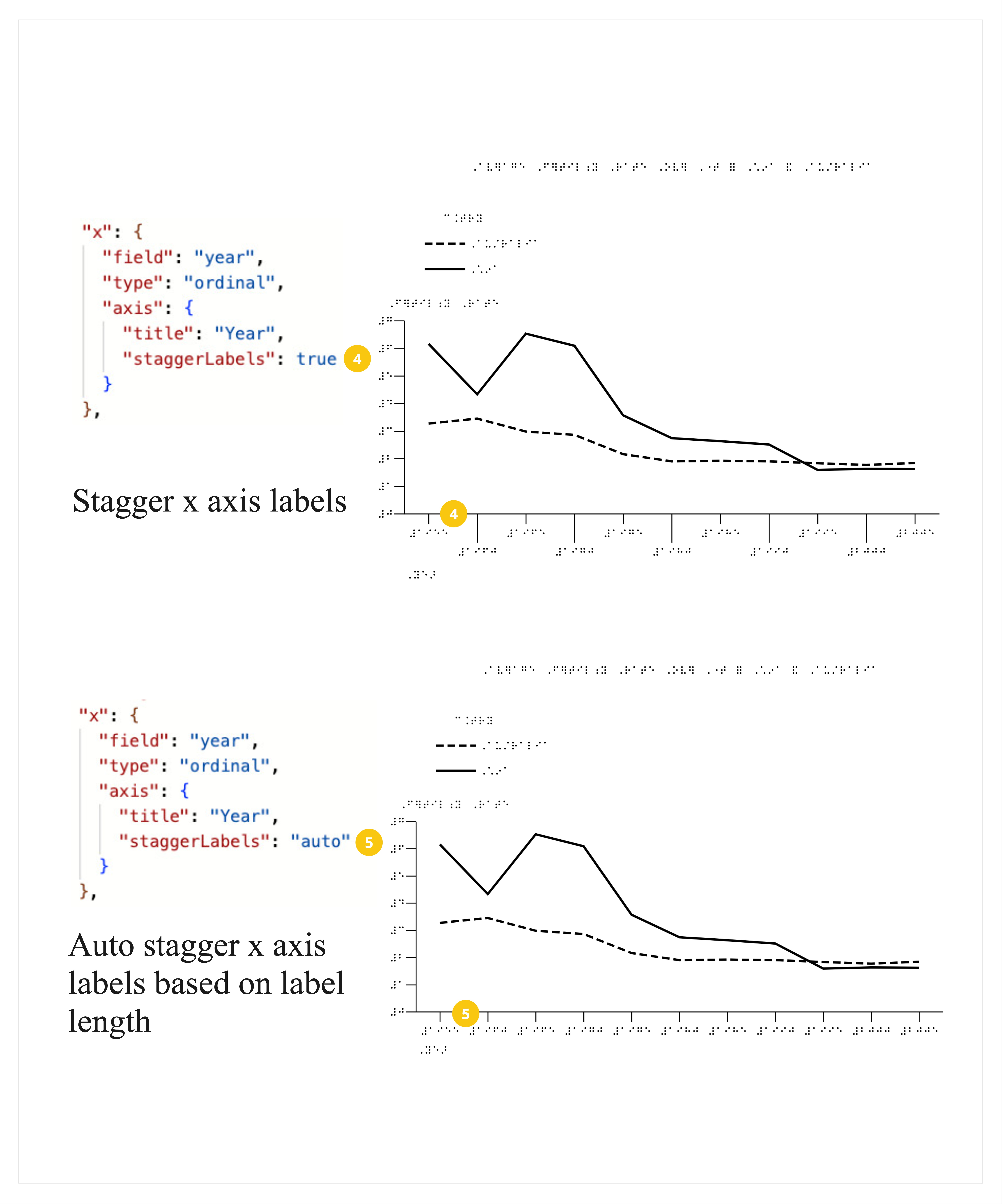}
        \Description{The image consists of two sections: the top chart uses `staggerLabels: true`, and the bottom chart uses `staggerLabels: "auto"`. Each section includes a JSON code snippet on the left and the corresponding tactile-rendered line chart on the right.
                Top Section: Stagger x-axis labels (`staggerLabels: true`)
        - JSON Code: The JSON code snippet defines the x-axis with the `"field"` set to `"year"`, `"type"` as `"ordinal"`, and `"staggerLabels": true`. This property ensures that x-axis labels are staggered for better spacing. This coerces the staggering of the x-axis labels, even if they are short enough to fit on the same line. 
        - Tactile Chart: The chart displays two lines:
            - A solid line representing one dataset.
            - A dashed line representing another dataset.
        - The x-axis labels are staggered, alternating positions above and below the axis to prevent overlap. Labels represent ordinal years (e.g., 1955, 1960, 1965).
        - The y-axis has braille labels indicating numeric values.
        - A legend at the top provides tactile descriptions of the two line styles (solid and dashed).
        Yellow marker number 4 is displayed next to the code and on the x-axis near the origin. 
        Bottom Section: Auto-stagger x-axis labels (`staggerLabels: "auto"`)
        - JSON Code: The JSON code snippet introduces `"staggerLabels": "auto"`, which dynamically staggers x-axis labels when it exceeds certain thresholds. based on their length to ensure legibility.
        - Tactile Chart:
          - The chart displays the same two lines (solid and dashed) as the top section.
          - The x-axis labels are not staggered.
          - Braille labels on the y-axis and a tactile legend remain consistent with the top chart.
        Yellow marker 5 is shown next to the code snippet and on the x-axis near the origin. 
             }
        \caption{Example of dynamically staggering labels based on their length.}
        \label{fig:example-stagger}
    \end{subfigure}
    \hfill
    \begin{subfigure}[t]{0.44\textwidth}
        \centering
        \includegraphics[width=\textwidth]{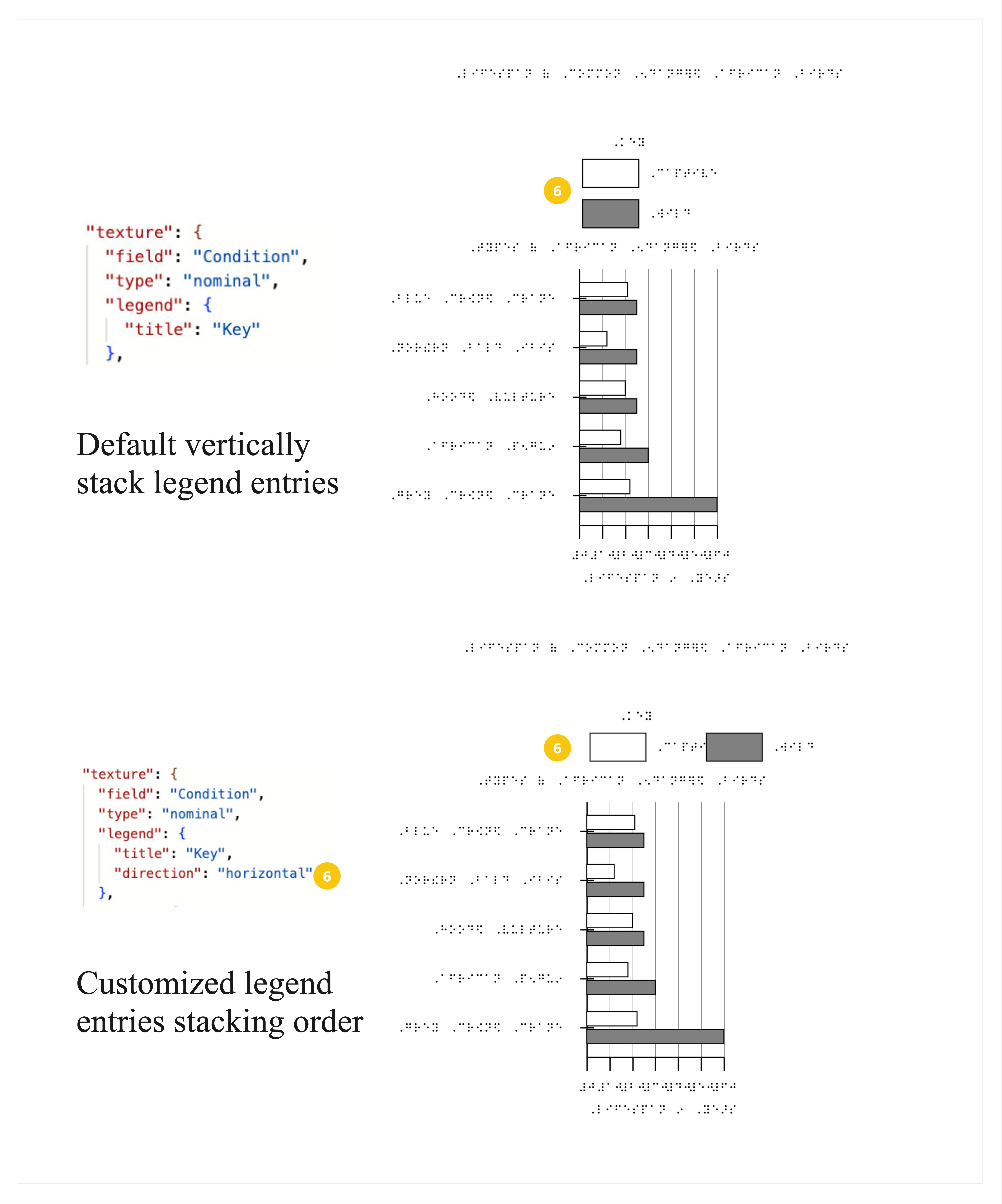}
        \Description{The image compares two tactile bar charts that demonstrate different legend orientations: vertically stacked legend entries (default) and horizontally aligned legend entries (customized). Each section includes a JSON code snippet on the left and the corresponding tactile-rendered bar chart on the right.
                Top Section: Default Vertically Stacked Legend Entries
        - JSON Code: The JSON snippet defines the `texture` field mapped to `"Condition"`, with a `"type": "nominal"`. The legend is titled `"Key"` with no additional configuration for direction, defaulting to a vertical stack.
        - Tactile Chart: displays data with segments in each bar encoded by different tactile textures representing `"Captive"` and `"Wild"` conditions.
            - The legend is positioned above the chart, with entries stacked vertically:
            - The first entry, an empty fill texture, represents `"Captive"`.
            - The second entry, a solid texture, represents `"Wild"`.
        - Each entry is annotated with braille labels.
        - Braille text labels the x-axis (e.g., categories such as bird species) and the y-axis (e.g., lifespan in years).
                Bottom Section: Customized Horizontal Legend Entries
        - JSON Code: The JSON snippet adds a `"direction": "horizontal"` property to the legend configuration, aligning entries horizontally.
        - Tactile Chart:
          - The tactile bar chart is identical to the top section in terms of data representation.
          - The legend is positioned above the chart, with entries aligned horizontally:
            - The first entry, an empty fill texture, represents `"Captive"`.
            - The second entry, a solid texture, represents `"Wild"`.
          - Braille labels annotate each legend entry.
          - Braille text labels the x-axis and y-axis, as in the top chart.}
        \caption{Comparison of default vertically stacked and customized horizontally aligned legend entries in tactile bar charts.}
        \label{fig:example-legend}
    \end{subfigure}
    \caption{Examples of tactile charts generated using TVL, showcasing both default and customized configurations for textures, gridlines, staggered labels, and legends.}
\end{figure*}

\paragraph{Positioning.} By default, the chart title is centered at the top of the page. This placement allows readers to quickly identify the chart's subject and orientation, as the title is typically the first element sought by blind users. If there is a legend, then the legend is positioned directly below the title, with each legend entry positioned vertically on the same page to maintain continuity. This helps readers familiarize themselves with the different textures and line styles before exploring the data points. Legend entries, by default, are stacked vertically to provide a straightforward, predictable reading path (\autoref{fig:example-legend}). Although placing the legend above the chart takes up space, it enables users to anticipate the chart content and recognize textures as they encounter them \cite{goncuTactileChartGeneration2008, engelAnalysisTactileChart2017}. 

\paragraph{Spacing.} The general rule of thumb for spacing is that 1/8 inch between any two elements is required to perceive individual pieces of information \cite{brailleauthorityofnorthamericaGuidelinesStandardsTactile2022}. TVL spaces axis labels 1/8 inch from the tick marks or axis lines on the x-axis and y-axis and adds additional padding between the axes and the marks in the chart. The system also adds spacing equivalent to 1 or 2 braille cell heights after the chart title and the x and y-axis titles. 
This spacing ensures that the tactile elements are not too densely packed, allowing users to differentiate between components. TVL sets the width of the chart based on these spacing specifications.  

\paragraph{Alignment.} By default, the system left-aligns the chart to the left of the plotting area to make it easier for readers to locate information quickly by systematically scanning from left to right. The x-axis title is left-aligned with the left-most x-axis label. We center the labels within the width of the bar or set of bars as recommended by the BANA guidelines \cite{brailleauthorityofnorthamericaGuidelinesStandardsTactile2022}. Similarly, the y-axis title is left-aligned with the y-axis labels, while the labels are center-aligned with the corresponding tick marks.

\subsection{Design Rationale}

Tactile Vega-Lite leverages \textit{smart defaults} to handle tedious formatting tasks, maintain consistency, and accelerate chart creation. By pre-configuring layout settings---including alignment, spacing, and positioning rules---TVL ensures that charts adhere to tactile design guidelines (DC1). These defaults eliminate the need for designers to manually adjust each element, speeding up the design process and reducing the cognitive load of applying complex guidelines. TVL smart defaults configure the navigational aids---such as grid lines, axis line width, and axis ticks---according to best practices for information hierarchy, ensuring consistency across multiple charts. 

In addition to being guideline-compliant, TVL’s default templates are reusable, allowing designers to easily apply consistent designs across different projects by loading the same TVL specification with new data. These reusable templates minimize manual adjustment while allowing designers to override default settings, ensuring that both efficiency and customization are balanced (DC1, DC2).

While smart defaults streamline the process, TVL recognizes that the needs of tactile readers are diverse, and designers often have the best knowledge of how to meet those needs. The system allows for complete customization of chart elements, enabling designers to adjust sizes, line styles, textures, and layout to suit specific contexts and users (DC2). This flexibility empowers designers to create tailored solutions while still benefiting from the efficiency of default settings. By combining smart defaults with customization, TVL strikes a balance between simplifying the design process and providing the control necessary to accommodate unique user requirements, addressing the trade-off between complexity and control (DC2).

TVL offers affordances for tactile design-specific needs with predefined styles for grid lines, textures, and line styles, allowing designers to quickly prototype charts without needing to create tactile assets manually. The system integrates braille translation directly, eliminating the need for designers to switch between multiple platforms like Duxbury and Illustrator. This built-in functionality starts with data and reduces the inefficiencies of moving across platforms, ensuring that tactile charts are produced efficiently and with fewer errors. By minimizing platform-switching and automating key steps, TVL enhances productivity and ensures designers can focus on higher-level decisions, solving key workflow issues specific to tactile chart design (DC3).

\subsection{Limitations}

Our implementation of TVL has several limitations. First, TVL currently only supports a subset of chart types supported by Vega-Lite. As we discuss in \autoref{sec:example-gallery}, we have prioritized the most common tactile chart types. However, it is also unclear for many other chart types what best practices are for translating them.
For example, area charts are not mentioned in the BANA guidelines, so we have not prioritized them in our prototype TVL compiler.
Second, TVL assumes a screen-based rendering model, limiting its ability to precisely model the size of the physically produced end result.
Tactile charts rely heavily on precise measurements for usability, but precise pixel-to-physical size calculations are dependent on hardware parameters.
For similar reasons, TVL does not currently provide machine-specific previews (e.g. for specific embossers).
Consequently, it might be difficult for users to anticipate how charts will look when produced on their devices. This lack of feedback can impact the quality of the final tactile outputs.

\clearpage
\section{Evaluation: Example Gallery}
\label{sec:example-gallery}

\begin{figure*}
    \includegraphics[width=0.86\textwidth]{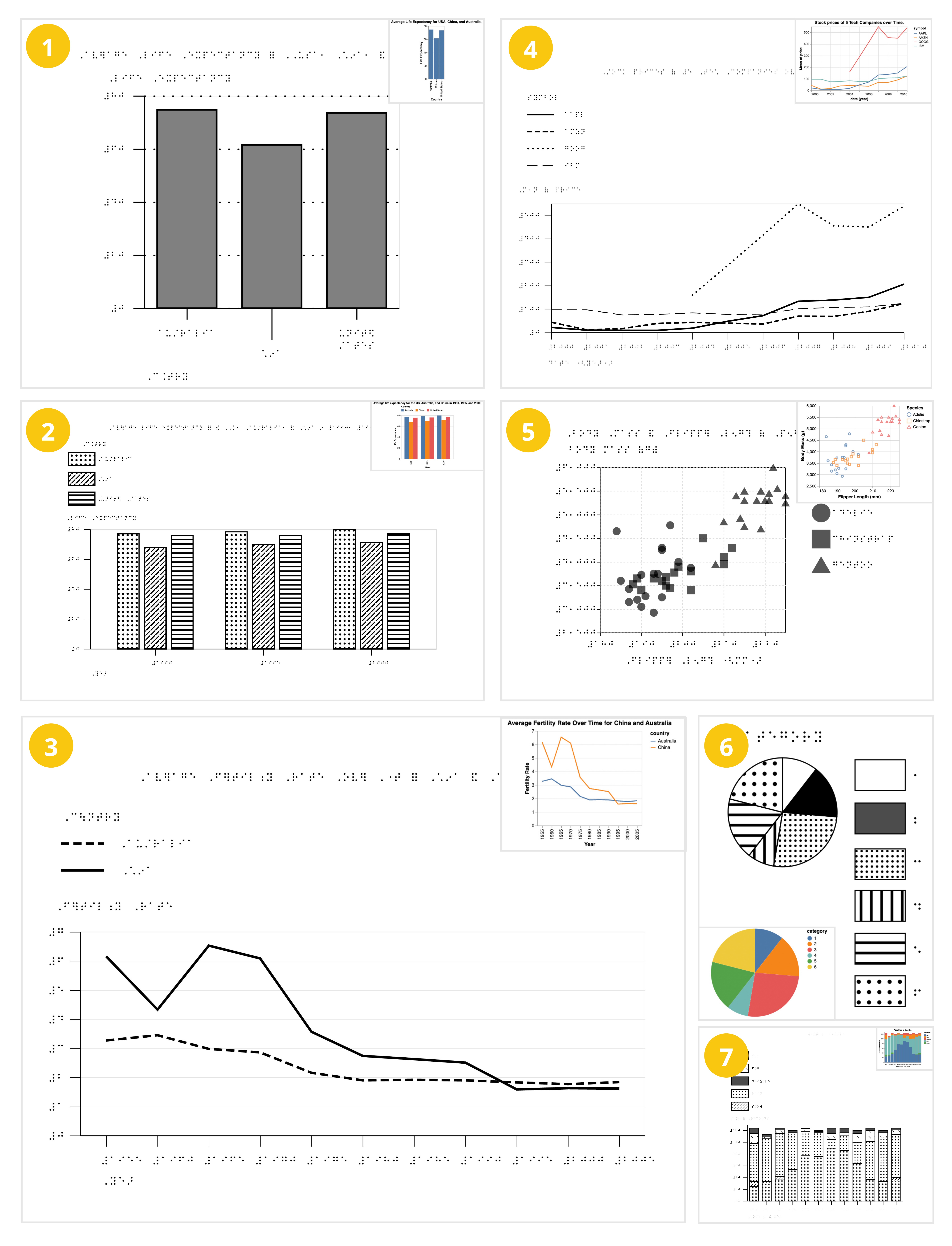}
    \Description{The image displays a gallery of tactile chart examples arranged in a grid with seven numbered panels, each showcasing a different type of tactile chart alongside its corresponding visual chart. From left to right, top to bottom: 
        1. Simple Bar Chart 
    - Content: the chart is titled "average life expectancy for USA, China and Australia". Y-axis ranges from 0 to 80 at a 20 interval. X-axis from left and right is Australia, China and United States. X-axis title is "Country" and y-axis title is "Life Expectancy".  
    - The bar chart has three vertical bars representing average life expectancy. The chart includes braille labels on the x- and y-axes, dotted tactile grid lines running horizontally across, and solid gray bar fills. 
    - The visual counterpart in the upper right corner shows a color-coded bar chart with labeled axes. The bars are in blue. X-axis labels and y-axis title are rotated 90 degrees counterclockwise. 
    2. Grouped Bar Chart
    - Content: the chart is titled "average life expectancy for USA, Australia and China". Y-axis ranges from 0 to 80 at a 20 interval. X-axis from left and right is 1990, 1995, 2000. There are three bars at each year, in the order of Australia, China and United States. X-axis title is "Year" and y-axis title is "Life Expectancy".  
    - A tactile grouped bar chart with multiple sets of bars distinguished by fill patterns such as densely dotted fill(Australia), diagonal left lines fill (China) and horizontal line fill(United States). Braille labels annotate the axes, and a tactile legend explains the patterns positioned right below the 
    - The visual chart in the upper right corner uses color to represent the same grouped data. Blue is Australia, orange is China, red is United States. 
    3. Dual Series Line Chart
    - Content: the chart is titled "average fertility rate over time for China and Australia". Y-axis ranges from 1 to 7 at 1 unit interval. X-axis from left and right from 1955 to 2005 at 5-year intervals. There are two lines one for China in orange and one for Australia in blue. X-axis title is "Year" and y-axis title is "Fertility Rate".  
    - A tactile line chart with two distinct lines: one solid (China)and one dashed (Australia). The chart includes braille labels on the axes, a legend with line styles, and horizontal solid gridlines. 
    - The visual chart in the upper right corner uses color-coded lines to represent the same trends.
       4. Multi-Series Line Chart
    - Content: the chart is titled "Stock prices of 5 techn companies over time". Y-axis ranges from 0 to 500 at 100 unit interval. X-axis from left and right from 2000 to 2010 at 1 year interval. There are four lines: red for google, cyan for IBM, blue for Apple, orange for Amazon. X-axis title is "date" in year and y-axis title is "Mean of price".  Google line starts at 2004 and trends upward significantly to 2010. Other companies starts at 2000 and show a gradual upward trend.
    - A tactile line chart with multiple lines, each differentiated by a unique line style. Dotted line for Google, solid line for Apple, long dashed line for IBM and a dashed line for Amazon. The tactile chart includes braille axis labels and a legend for line styles positioned below the chart title and above the chart area. 
    - The visual chart in the upper right corner uses colors to represent each line.
       5. Scatter Plot
    - Content: The chart is a scatter plot of flipper length and body mass of three species of penguins. Y-axis, titled "body mass(g)", ranges from 2500 to 6000 at 500 unit interval. X-axis, titled "flipper length(mm)", from left and right from 180 to 220 at 10 mm interval. There three types of penguins: Adelie (Blue circle), Chinstrap (orange square) and Gentoo (red triangle). We see a positive trend between flipper length and body mass, body mass increases as flipper length increases. 
    - A tactile scatter plot with points represented by different shapes (e.g., circles, triangles, squares) to encode categories. The axes have braille labels, and a tactile legend explains the shapes. 
    - The visual chart in the upper right corner uses colored markers for the same data.
    6. Pie Chart
    - Content: toy dataset of a pie chart with 6 categories, each colored differently. Category 1 is blue, 2 is orange, 3 is red, 4 is cyan, 5 is green and 6 is yellow. 
    - A tactile pie chart with segments distinguished by fill patterns category 1 is empty fill, 2 is solid fill, 3 densely dotted, 4 is vertical lines, 5 is horizontal lines, 6 is a dotted fill. A tactile legend explains the patterns, and braille labels annotate the chart. 
    - The visual chart in the lower right corner uses colors for each segment.
    7. Stacked Bar Chart
    - Content: the chart is titled "Weather in Seattle". The Y-axis titled "Count of Records" ranges from 0 to 120 at 20-unit intervals. The X-axis is titled "Month of the Year" from the left and right from Jan to Dec at 1-month intervals. There are five types of weather: sunny(in blue), fog(in orange), drizzle(in red), rain(in cyan), and snow(in green). The bars are stacked vertically, with sunny conditions at the bottom, then snow, rain, fog, and drizzle.
    There is higher percentage of sunny days during summer months. 
    - A tactile stacked bar chart with segments in each bar differentiated by textures such as diagonal lines and dots. A tactile legend explains the patterns, and braille labels annotate the axes. The visual chart in the corner uses colors for the stacked segments.}
    \caption{Gallery of tactile chart examples demonstrating various chart types, including bar charts, line charts, scatter plots, pie charts, and stacked bar charts. Each tactile chart is accompanied by its visual counterpart, showcasing how texture-based encodings, braille annotations, and tactile legends make data tactually accessible.}
    \label{fig:example_gallery}
\end{figure*}

The Tactile Vega-Lite (TVL) example gallery showcases a diverse range of charts and graphs, bar charts, line charts, pie charts, and stacked and grouped bar charts, highlighting the versatility and adaptability of the tool for presenting complex data in tactile-friendly formats.
To understand what chart types were most important to express, we collected 49 tactile chart examples from various sources, including guidelines, research papers, institutional libraries, and industry standards \cite{barthTactileGraphicsGuidebook1982, edmanTactileGraphics1992, rosenbergTouchingNews2024, TactileGraphicsImage2019, brailleauthorityofnorthamericaGuidelinesStandardsTactile2022,
americanprintinghousefortheblindinc.GoodTactileGraphic1998}. Our categorization revealed a dominant preference for bar charts (both grouped and stacked), line charts, and pie charts. As a result, we prioritized these simpler, more commonly used charts in the TVL example gallery.
However, TVL cannot easily express charts not within base Vega-Lite's expressive gamut.
For example, network visualizations are outside of TVL's scope.

The simple bar chart (\autoref{fig:example_gallery}.1) illustrates how x-axis labels are staggered for better tactile readability, as well as the \texttt{dottedGrid} configuration. The grouped and stacked bar charts (\autoref{fig:example_gallery}.2) and (\autoref{fig:example_gallery}.6), along with the pie chart (\autoref{fig:example_gallery}.7), showcase tactually distinct textures and with different legend positions, allowing readers to locate and familiarize themselves with the legend before exploring the chart itself. The two-series line chart (\autoref{fig:example_gallery}.3) demonstrates the hierarchical relationship between encoded lines and navigational aids, with the encoded lines representing data being the most prominent. The multi-series line chart (\autoref{fig:example_gallery}.4) highlights the variety of line styles available. Lastly, the scatterplot (\autoref{fig:example_gallery}.5) demonstrates different tactile shape encodings available.

\begin{figure*}[ht!]
    \includegraphics[width = \textwidth]{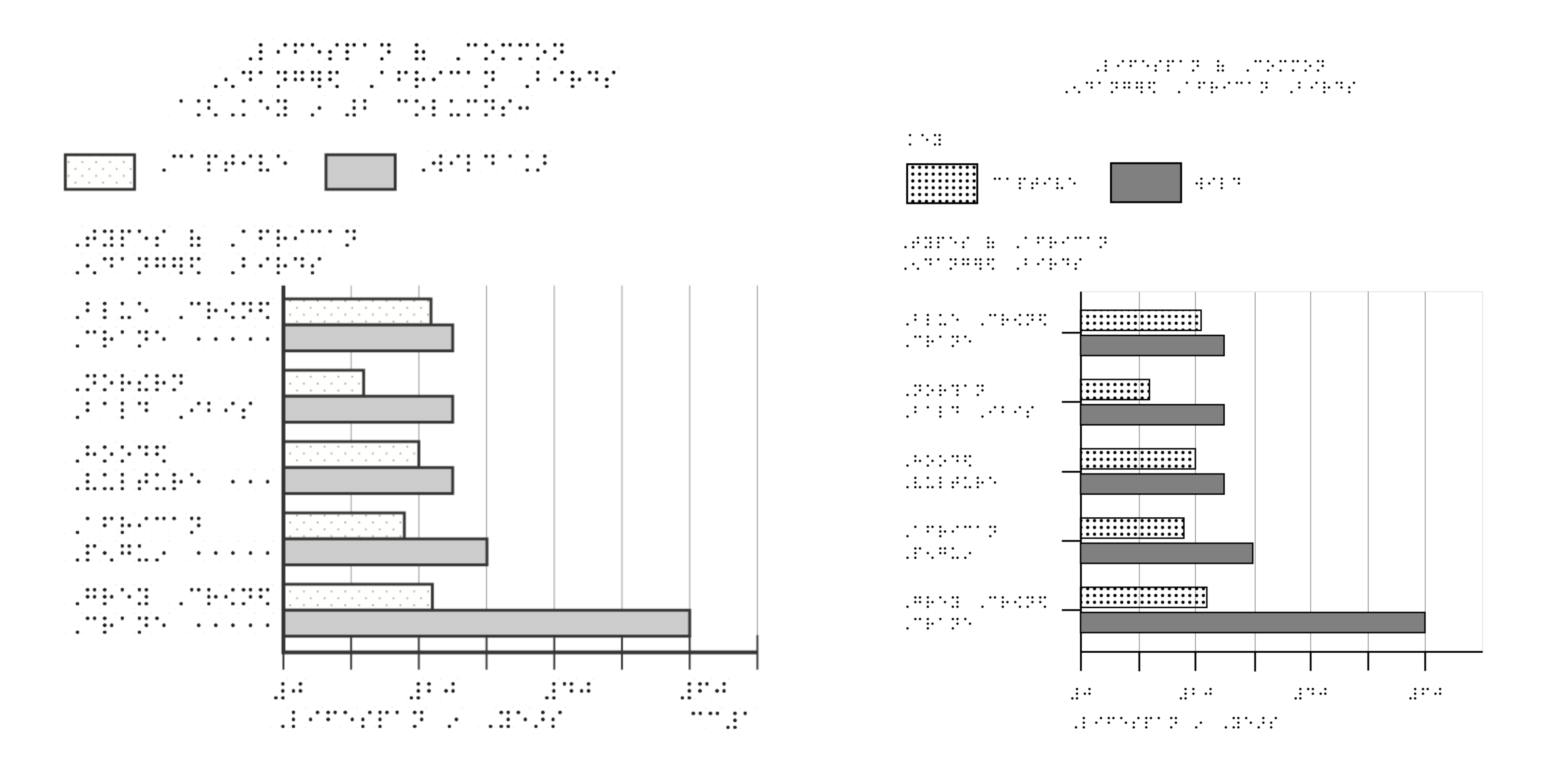}
    \Description{This figure consists of two tactile bar charts with corresponding braille text and legends, each designed to encode data using textures and braille labels. 
    Left Chart - Original Version of the tactile chart produced by an expert:
    The chart title, in braille, is positioned at the top and reads "Lifespan of Common Endangered African Birds." The title is broken into multiple lines. 
    A legend is included above the chart area. It uses two tactile fills to differentiate between conditions: "Captive," represented by a dotted fill, and "Wild," represented by a solid gray fill, both annotated with braille labels. The two legend entries are side by side on the same row. 
    The x-axis, located at the bottom, is labeled in braille with lifespan ranges, including values such as 0, 10, 20, 30, 40, 50, and 60 years.
    The y-axis, on the left, lists bird names in braille, such as "Blue Crowned Crane," "Northern Bald Ibis," and "Hooded Vulture."
    Horizontal bars represent the average lifespan of each bird species, with tactile textures corresponding to the legend. Bars representing animals living in the wild are shown with a solid gray fill. Bars representing animals living in captive are shown with a dotted pattern. 
    Bars are aligned with their respective bird names and scaled according to lifespan values.
    Right Chart (Replicated Version using tactile vega-lite):
    - The chart title, in braille, is identical to the original: "Lifespan of Common Endangered African Birds."
    - The legend, positioned above the chart, uses the same textures as the original: dotted fill for "Captive" and solid gray fill for "Wild," annotated with braille text. There is an additional legend title in braille that says "key"
    - The x-axis and y-axis follow the same layout as the original, with braille labels for lifespan ranges and bird names, respectively. The difference is that the tick marks are added. 
    - Bars are horizontally aligned with bird names and scaled to the same lifespan values as the original chart. The tactile textures for "Captive" and "Wild" conditions are consistent with the original version. The dotted texture used is a bit denser and not on an offset. }
    \caption{Comparison of original and replicated tactile bar charts representing the lifespan of common endangered African birds. Left graphic courtesy of the Media and Accessible Design Lab at LightHouse, San Francisco, and An Intervention to Provide Youth with Visual Impairments with Strategies to Access Graphical Information in Math Word Problems project funded by the Institute of Education Sciences (R324A160154).}
    \label{fig:replicate}
\end{figure*}

To show that TVL is expressive enough to support real-world use cases, we replicated an example chart created by an expert tactile designer (\autoref{fig:replicate}).
In this case, the smart defaults for a grouped bar chart in TVL were able to mostly approximate the original design.
We made a series of customizations so that our chart more closely matches the real-world example.
These included manually choosing the two textures from the TVL palette that most closely matched the original, manually adding line breaks to the chart title, and adjusting the width of the chart to create similar spacing.
This exercise demonstrates that we could rapidly author a chart in TVL that closely approximates the layout and texture choices made by a professional designer.

\section{Evaluation: User Study}
\label{sec:user-study}

\subsection{Methods}

\paragraph{Participants.} We recruited 12 participants, including four tactile graphics designers, three teachers of students with visual impairment  (TVIs), and five braille transcribers. All participants indicated they have considerable or expert-level experience with creating tactile graphics. 66.7\% (n=8) participants self-rated as extremely confident in their ability to create tactile graphics. 41.7\% (n=5) participants had more than 10 years of experience, 41.7\% (n=5) participants had 5-10 years of experience. 41.7\% (n=5) participants created tactile graphics every day, 25\% (n=3) participants created tactile graphics once or twice a week. Adobe Illustrator and physical objects were the most frequently used tools for creating tactile graphics, followed by CorelDraw and Tiger Designer Suite. 91.7\% (n=11) participants used embossers to produce tactile graphics, and 75\% (n=9) participants used swell paper. Participants' other production methods included 3D printing, hand-drawn methods, UV printing, vacuum forming, and collages. Among the responses, participants identified the time-consuming nature of creating tactile graphics as the most significant challenge, followed by technical difficulties and lack of training and knowledge. In the examples shared by participants, most participants have made at least one of the following chart types: bar, pie, line chart. 

\paragraph{Study setup.} We interviewed each of our 12 participants for 60 minutes. We began the session with open-ended interview questions centered around the participant's current tactile graphics creation process. We then presented participants with two example designs in the TVL editor. Participants interacted with each design for about 20 minutes. The designs utilized TVL's smart defaults to implement guidelines, recommendations, and known best practices. Participants first critiqued and evaluated the default charts created by the system. Participants then made their desired adjustments to the default charts to meet their standards or requirements. When participants had little or no critique, we prompted them to explore the chart with a list of tasks designed to evaluate the predefined styles. As participants modified the chart, we asked them questions to understand their thought process. We then conducted a closing interview and asked participants to complete a Likert scale survey to evaluate the system and its defaults. 

\paragraph{Example charts.} Both examples used the \texttt{gapminder.json} dataset from the Vega dataset repository \cite{roslingBestStatsYou2006}. The dataset includes the fertility rate and life expectancy of countries around the world from 1955 to 2005, at a 5-year interval. The first example was a bar chart that displays the average life expectancy for the USA, China, and Australia (\autoref{fig:example_gallery}.1). The second example was a multi-series line chart showing China and Australia's average fertility rate from 1955 to 2005 (\autoref{fig:example_gallery}.3).

\paragraph{Analysis.} We analyzed user testing results by reviewing video recordings, transcripts, and notes taken during the 12 interview sessions. We used an open coding approach to annotate transcripts, following a grounded theory method. We then did another pass to categorize these codes into broader themes, such as feedback on the tool’s functionality, the default specification, customization capabilities, chart-specific concerns, and suggestions for improvement.

\paragraph{Motivation.} Our study prioritized an exploratory focus to understand the diverse practices and challenges faced by tactile graphics creators.
Given the variability in workflows, tools, and priorities among professionals like educators, designers, and braille transcribers, we sought to observe how experts interacted with TVL's features to critique and customize default charts, revealing their design rationale and iterative processes. 
This approach allowed us to gather rich, context-specific insights while working within the practical constraints of a 60-minute session.

\subsection{Survey Results}
We designed a Likert survey to evaluate the system's default choices and understand the usefulness of predefined styles. Results are shown in (\autoref{tab:quant-results}). 

\begin{figure*}
    \includegraphics[width=\textwidth]{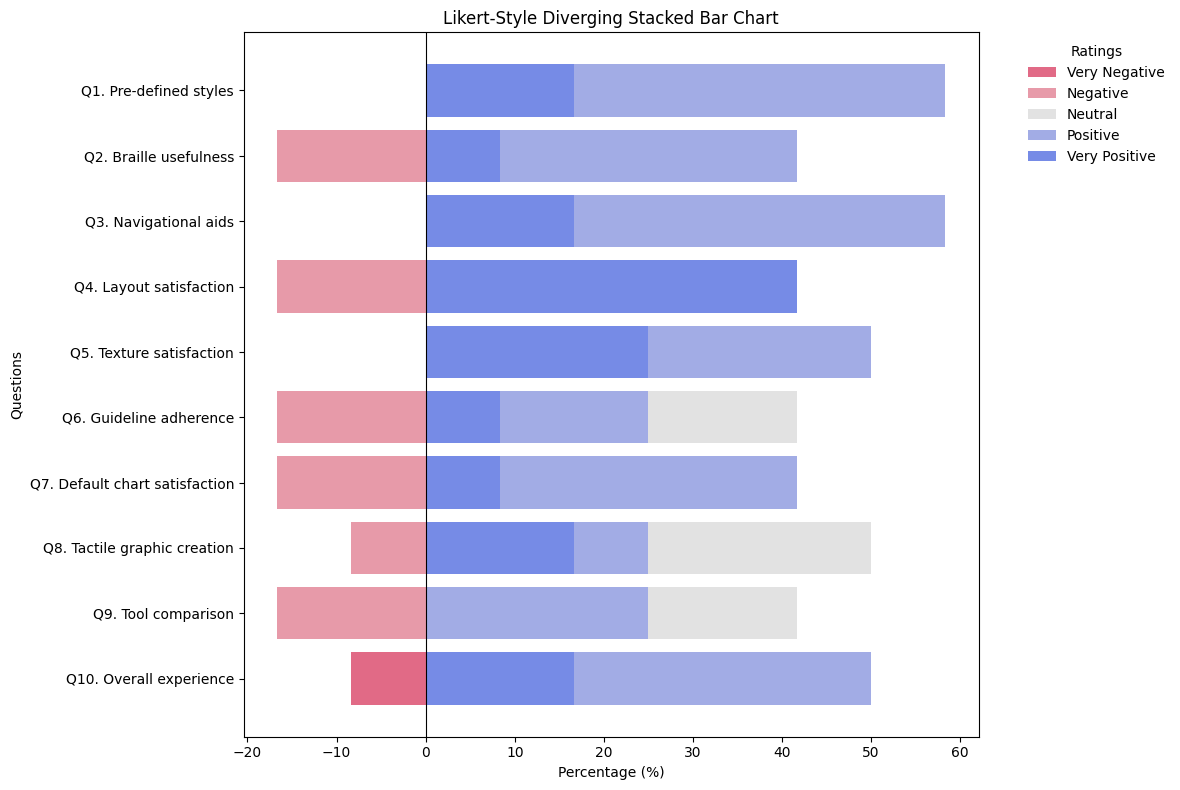}
    \Description{The image shows a Likert-style diverging stacked bar chart that visualizes user satisfaction ratings across 12 different survey questions. The x-axis represents percentage values ranging from -20\% to +60\%, with a vertical line at 0\% dividing negative and positive responses. The y-axis lists the survey questions.
    Chart Details:
    1. Title: The chart is titled "Likert-Style Diverging Stacked Bar Chart," positioned at the top.
    2. X-axis: Labeled "Percentage (\%)" with tick marks at intervals of -20, -10, 0, 10, 20, 30, 40, 50, and 60.
    3. Y-axis: Labeled "Questions," with the following survey questions listed:
   - Q1. Predefined styles
   - Q2. Braille usefulness
   - Q3. Navigational aids
   - Q4. Layout satisfaction
   - Q5. Texture satisfaction
   - Q6. Guideline adherence
   - Q7. Default chart satisfaction
   - Q8. Tactile graphic creation
   - Q9. Tool comparison
   - Q10. Overall experience
    4. Bars:
   - Each question is represented by a horizontal stacked bar, diverging from the center at 0\%.
   - The bars are color-coded to represent user ratings:
     - Red (left side): Very Negative and Negative ratings.
     - Gray (center): Neutral ratings.
     - Blue (right side): Positive and Very Positive ratings.
   - The width of each section reflects the percentage of respondents selecting that rating.
    5. Legend: Positioned on the right, the legend indicates the color coding:
   - Red: Very Negative, Negative.
   - Gray: Neutral.
   - Blue: Positive, Very Positive.
The data for this chart can be found in Table 1 in the paper. 
}
    \caption{Study results show positive responses for styles(Q1), texture designs(Q5), navigational design(Q3), and mixed feedback for other categories like guideline adherence (Q6) and tool comparability(Q9).}
    \label{fig:user_study_res}
\end{figure*}

\begin{table*}
  \renewcommand{\arraystretch}{1.2} %
  \caption{Rating scores for different aspects of the default tactile charts and customization experience on a five-point Likert scale where {1} = Not useful or satisfied at all and {5} = Extremely useful or satisfied. Median scores are shown in \textbf{bold}, averages in brackets [], and standard deviations in parentheses ().}
  \label{tab:quant-results}
  \begin{tabular*}{\textwidth}{@{\extracolsep{\fill}} p{0.75\textwidth} p{0.15\textwidth}}
    \toprule
    Survey Questions & Score \\
    \midrule
    Q1. How would you describe the predefined line styles in the prototype? & \textbf{4} [3.9] (0.67) \\
    Q2. How satisfied are you with the predefined textures in the prototype? & \textbf{4} [4] (0.74) \\
    Q3. How useful was the braille provided by the prototype? & \textbf{3.5} [3.4] (0.90) \\
    Q4. How useful were the navigational aids (grid lines, grid line styles, etc) in the prototype? & \textbf{4} [3.9] (0.67) \\
    Q5. How satisfied are you with the layout of the prototype? & \textbf{4} [4] (1.1) \\
    Q6. How well do you think the output of the prototype adheres to guidelines? & \textbf{3} [3.1] (1.08) \\
    Q7. How satisfied are you with the default chart produced by the prototype? & \textbf{3.5} [3.3] (1.14) \\
    Q8. Are you able to create the tactile graphics that you wanted to create using our prototype? & \textbf{3} [3.5] (0.90) \\
    Q9. How do the prototype's capabilities compare to the tools that you are using? & \textbf{3} [2.8] (1.07) \\
    Q10. How would you rate your overall experience with the prototype?  & \textbf{4} [3.7] (1.07) \\
    \bottomrule
  \end{tabular*}
\end{table*}

Responses suggested that participants highly appreciated predefined line styles (Q1) and textures (Q2), as they offer consistency and reduce design time.
Our interviews offered further context to these responses.
For instance, P1 highlighted that pre-built textures are very similar to their existing workflow in that they have a texture sampler that they frequently refer to. Participants generally liked the default texture and line style choices and shared preferences for directional patterns (like \texttt{DiagonalLeft}, \texttt{DiagonalRight}) (P1, P7, P10).
Participants also particularly liked the density of the dotted textures: not too dense such that it feels like a smooth surface, but also not too loose (P2, P3). P1 liked being able to ``set differentiable tactile hierarchy for different types of lines.''

Similarly, participants thought that using grid lines (Q4) was beneficial for helping students track information and orienting readers (P12), particularly for younger kids (P10). Even though participants noted that grid lines could make the design more cluttered and tactually complex (P9), they would still use them for younger kids who often have ``a hard time bringing their hands horizontally across the page, needing some sort of horizontal grid lines'' (P8). This further illustrates that the reader's needs and abilities play a key role in these design decisions and, thus, the importance of customizability. Regarding layout (Q5), there was general agreement that spacing and alignment are critical to ensuring tactile legibility, with participants preferring well-aligned axis labels and sufficient spacing around key elements like braille and marks (P9, P10).

The relatively modest scores for braille usefulness (Q3) reflect the limitations of our initial implementation. Specifically, the text was converted to lowercase, leading to discrepancies with the visual representation in capitalization.
Certain braille rules were not fully implemented, such as when to include or omit numeric indicators, were not fully implemented.
These challenges highlight the complexities of braille translation.
While the current implementation provides a functional starting point, we aim to incorporate this feedback into future iterations of TVL.

Interestingly, the relatively middling score for adherence to guidelines (Q6, Q7) revealed differing expectations among participants.
In our design process, we did our best to implement TVL according to the guidelines.
However, the implementation we used in the user study was imperfect.
TVIs, who may prioritize practical usability over strict adherence to guidelines, were generally less concerned with whether every guideline was followed precisely. In contrast, guideline writers, who possess a strong commitment to established standards and a high familiarity with those guidelines, were more likely to scrutinize the output charts rigorously for adherence and identify more issues.
Specific issues noted by participants included tick marks not straddling the axis and the lack of white space around the intersection of the line charts. 

Q8's score reflects the challenges participants faced with the prototype’s specification-based interface. Several participants (P5, P6, P7, P8, P12) commented that introducing a user-friendly graphical interface would improve ease of use, particularly for users less familiar with programmatic design. We plan to incorporate this feedback in future versions of TVL.

Participants' varied perceptions of how TVL compares to existing tools (Q9) can be attributed to two primary factors. First, many participants, particularly educators, expressed reluctance to adopt new tools due to time constraints and a lack of resources. In certain areas, access to technology is limited, and introducing new tools requires state approval and additional training for educators, making the process difficult and slow. Second, professional tactile graphics designers who have developed a high level of proficiency with powerful design software like CorelDraw and Illustrator, often find these established tools more suitable for their advanced needs, which may have influenced their lower scores for the prototype.

\subsection{User Study Results}

\subsubsection{Designers break rules to meet individual reader needs.}
Adhering to the guidelines was important for our participants, especially when designing for broader audiences or educational purposes.
However, we found broad agreement that breaking from the guidelines is justified when it enhances functionality. 
Participants explained that they break format rules when working individually with students (P5) or when important information needs to be included, and the only way to do so effectively goes against the guidelines (P12).
P5 emphasized that ``guidelines are not rules'' and highlighted the flexibility needed to accommodate unique scenarios, stating, ``There’s nothing specifically set in stone that you can’t adjust.'' This flexibility is essential given the diverse experience levels of tactile graphics readers, with P2 noting that the effectiveness of a graphic can vary significantly depending on the reader’s familiarity and skill.

\subsubsection{Educators need charts efficiently, even at the expense of guidelines}
When making charts to address immediate needs, efficiency often takes priority over perfection. P8, a teacher, noted, ``I do my best to include the essential things… but would my spacing be perfect? No.'' This prioritization reflects the reality of balancing guideline adherence with the demands of lesson planning and grading. P10 confirmed that guidelines are often deprioritized in favor of focusing on student academics, stating, ``We don't use guidelines as much as we should.'' 

However, our educator participants also reflected on the broader implications of not consistently following guidelines, particularly in high-stakes contexts like standardized testing. P10 underscored the importance of guidelines for ensuring consistency in state testing, pointing out the potential challenges when students are suddenly exposed to strictly guideline-compliant materials after using less standardized resources. This highlights the importance of striking a balance between rapid authoring, flexibility, and maintaining standardization for critical applications like assessments.

For participants, these observations underscore the importance of tools like TVL to their work.
What often gets overlooked or deprioritized during a time crunch are the more manual and tedious aspects of tactile graphic design, such as ensuring proper spacing and consistent formatting. 
Participants appreciated that TVL automated these tedious details. P2, who has tried creating scripts for Adobe Illustrator to ``automate processes for standardization'',   commented that found ``it’s great to have [TVL] do all the formatting because that’s a huge time saver to make sure things space out.'' P2 felt that they could ``generate [charts] quickly'' without having to worry about implementing every low-level guideline. 

\subsubsection{Expert design critique reinforced the importance of customizability}
When we asked expert designers to give feedback on example default designs, there was room for reasonable disagreement between experts.
Prompted by the predefined line styles in TVL, participants debated ways of expressing hierarchy through line weight and styling, as well as the inclusion of grid lines. 
To express the least tactually significant grid lines, P5 and P10 used a solid line with hairline thickness for grid lines, expressing concern that readers might confuse \texttt{dottedGrid} with ``just random dots being embossed.''
However, P7 advocated for dotted grid lines, stating, ``I think they are better as the lightest thing.'' P11 suggested that grid lines could have different textures to make them fainter, further highlighting the diverse approaches that designers take to meet different readers’ needs.

The use of grid lines in a multi-series line chart was another point of contention. P7 would add grid lines for readers, just to make their lives a little easier, even though grid lines are not needed in this situation according to guidelines. P8 and P12 disagreed on whether to provide horizontal grid lines for value readout and vertical grid lines for tracing the year axis---P8 leaned towards only vertical grid lines, while P12 created grid lines for both axes.

These varying opinions emphasize that, as P5 pointed out, ``If you give designers the same data, you will get 20 different things.''
In other words, there is no single best chart design. This comment highlights the inherent variability in how readers perceive tactile charts and underscores the need for tools to support customization and flexibility.

\subsubsection{Instant feedback accelerates tactile design workflows. }
Participants appreciated the immediate feedback provided by the TVL editor and tactile renderer. P7 remarked, ``[TVL] creates instantly, there’s nothing else that does that,'' while P10 noted that they like it when they can ``alter [TVL spec] and see it in real-time. There’s no lag and having to wait and see how it’s gonna look.'' Additionally, P10 emphasized the advantage of simultaneous editing and visualization, noting that they could make edits as needed while having the original graph readily available for reference. Participants identified the ability to preview how designs would be embossed on different embossers as essential, although this presents challenges due to the proprietary nature of many embossing machines. Incorporating a ``send to printer'' feature, as suggested by P1, could significantly enhance TVL’s utility. This functionality would allow direct connections to output devices like embossers, facilitating seamless transitions from design to production and integrating the entire workflow into a streamlined system.

\subsubsection{Balancing tactile-first design with information equity is difficult and important. }
Participants revealed a push-and-pull between prioritizing tactile-specific design and ensuring that tactile charts convey the same information as their visual counterparts, particularly in maintaining the integrity of data interpretation. P7 highlighted the importance of replicating visual impressions, such as the dramatic drop in data, to ensure blind students receive the same contextual information as sighted students. This concern reflects the need for tactile charts to be as visually similar as possible in certain contexts while still accommodating the tactile medium’s unique requirements.

\section{Discussion and Future Work}

In this paper, we presented Tactile Vega-Lite (TVL), a system for rapidly prototyping tactile charts using extensions to the Vega-Lite grammar of graphics.
We motivated the design of TVL through formative interviews with seven expert tactile designers, surfacing rich insights about the challenges of existing tactile design workflows.
For instance, non-experts often struggle with the complexity of tactile guidelines, while experts are hindered by fragmented, time-consuming workflows that require extensive manual formatting.
We address these challenges through TVL's tactile-specific abstractions, which provide smart defaults that allow non-experts to quickly generate guideline-aligned tactile charts and experts to prototype and customize based on reader’s needs.
A user study with 12 participants found that TVL enhances both flexibility and consistency by automating tedious tasks such as adjusting spacing and translating braille, and accelerates design iterations with pre-defined textures and line styles.

\subsection{Tactile charts and multimodality}

Recent work in accessible data representations has underscored the importance of using multiple data representations together in a complementary fashion \cite{zongUmweltAccessibleStructured2024b, sharifUnderstandingScreenReaderUsers2021, sharifVoxLensMakingOnline2022, thompsonChartReaderAccessible2023}.
We found that many moments in our user studies support this idea.
For example, a recurring request from designers (P1, P10) was the ability to overlay visual representations on tactile charts. 
Participants referenced the importance of hybrid visual and tactile representations to low-vision users (P9).
In particular, one participant mentioned that she started introducing one of her low-vision students to tactile charts because they often get fatigued when only reading visually.
Another participant (P1) noted that it is hard to read braille visually, so low-vision users might also benefit from the presence of both text and braille.
These insights suggest that moving back and forth between multiple modalities can benefit users, so researchers should be attentive to ways of incorporating tactile charts into multimodal systems, as many have begun to do \cite{chunduryTactualPlotSpatializingData2023, bakerTactileGraphicsVoice2016a,10.1145/3613904.3642730}.

\subsection{Perceptual research on tactile information hierarchy}

In our design process and user studies, we noticed a relative lack of consensus in perceptual research on tactile graphics when it comes to information hierarchy. A key question is how different techniques for creating tactual distinctiveness---such as line styles, thickness, and placement---affects users' ability to understand information hierarchy. Comparing the relative perceptual effectiveness of these tactile features can provide insights into which designs are more easily distinguishable and effective in conveying complex data.
Users' strategies for tactile reading and navigation could also changes which approaches to tactile information hierarchy are most effective. Understanding these distinctions can guide the development of best practices for creating tactile data representations that ensure BLV readers are able to quickly discover and navigate chart elements.

\subsection{Designing tactile-first data representations}

Tactile charts serve an important purpose in making existing visualizations accessible.
This is important for helping blind and low-vision readers establish common ground with sighted readers of visualizations, especially in educational settings where students are learning how to use charts.
However, as a result, the goal of tactile chart designers has largely been to faithfully reproduce visual chart forms in the tactile modality.

In our design process and exploratory tests with blind readers, we sometimes felt that existing visual forms were not well suited to tactile representation.
For example, bar charts have a lot of empty interior space.
While they are helpful for making length comparisons visually, they provide less tactile signal.
Often there were other possible chart forms (like a dot plot with a size encoding) that seemed potentially more suited to the same data from a tactile perspective.
For blind designers creating tactile graphics directly (instead of trying to translate existing visualizations), future work has an opportunity to advance tactile data representation by designing and evaluating the effectiveness of tactile-first data representations.

\begin{acks}
This work was supported by NSF \#2341748.

For those who helped inform this project and who were study participants and have consented to be acknowledged by name, we thank them here in alphabetical order by last name: Rich Caloggero, Christina Conroy, Chancey Fleet, Stacy Fontenot, John Gardner, Steve Landau, Eleanor Mayes, Jessica McDowell, Lisa Marie Mejia, Josh Miele, Janet Shirley Milbury, Anitha Muthukumaran, Ken Perry, Robin Ramos, Naomi Rosenberg, Penny Rosenblum, Yue-Ting Siu. 
\end{acks}
\bibliographystyle{ACM-Reference-Format}

\begin{thebibliography}{58}


\ifx \showCODEN    \undefined \def \showCODEN     #1{\unskip}     \fi
\ifx \showDOI      \undefined \def \showDOI       #1{#1}\fi
\ifx \showISBNx    \undefined \def \showISBNx     #1{\unskip}     \fi
\ifx \showISBNxiii \undefined \def \showISBNxiii  #1{\unskip}     \fi
\ifx \showISSN     \undefined \def \showISSN      #1{\unskip}     \fi
\ifx \showLCCN     \undefined \def \showLCCN      #1{\unskip}     \fi
\ifx \shownote     \undefined \def \shownote      #1{#1}          \fi
\ifx \showarticletitle \undefined \def \showarticletitle #1{#1}   \fi
\ifx \showURL      \undefined \def \showURL       {\relax}        \fi
\providecommand\bibfield[2]{#2}
\providecommand\bibinfo[2]{#2}
\providecommand\natexlab[1]{#1}
\providecommand\showeprint[2][]{arXiv:#2}

\bibitem[~(2019)]%
        {TactileGraphicsImage2019}
\bibfield{author}{\bibinfo{person}{American Printing~House}.}
  \bibinfo{year}{2019}\natexlab{}.
\newblock \bibinfo{title}{The {Tactile} {Graphics} {Image} {Library}: {Helping}
  {Students} {Succeed}}.
\newblock
\newblock
\urldef\tempurl%
\url{https://www.aph.org/the-tactile-graphics-image-library-helping-students-succeed/}
\showURL{%
\tempurl}


\bibitem[Aldrich et~al\mbox{.}(2003)]%
        {aldrichFirstStepsModel2003}
\bibfield{author}{\bibinfo{person}{Frances Aldrich}, \bibinfo{person}{Linda
  Sheppard}, {and} \bibinfo{person}{Yvonne Hindle}.}
  \bibinfo{year}{2003}\natexlab{}.
\newblock \showarticletitle{First {Steps} {Towards} a {Model} of {Tactile}
  {Graphicacy}}.
\newblock \bibinfo{journal}{\emph{Cartographic Journal}} \bibinfo{volume}{40},
  \bibinfo{number}{3} (\bibinfo{date}{Dec.} \bibinfo{year}{2003}),
  \bibinfo{pages}{283--287}.
\newblock
\showISSN{00087041}
\urldef\tempurl%
\url{https://doi.org/10.1179/000870403225013014}
\showDOI{\tempurl}
\newblock
\shownote{Publisher: Taylor \& Francis Ltd}.


\bibitem[Aldrich and Parkin(1987)]%
        {aldrichTangibleLineGraphs1987}
\bibfield{author}{\bibinfo{person}{Frances~K. Aldrich} {and}
  \bibinfo{person}{Alan~J. Parkin}.} \bibinfo{year}{1987}\natexlab{}.
\newblock \showarticletitle{Tangible {Line} {Graphs}: {An} {Experimental}
  {Investigation} of {Three} {Formats} {Using} {Capsule} {Paper}}.
\newblock \bibinfo{journal}{\emph{Human Factors}} \bibinfo{volume}{29},
  \bibinfo{number}{3} (\bibinfo{date}{June} \bibinfo{year}{1987}),
  \bibinfo{pages}{301--309}.
\newblock
\showISSN{0018-7208}
\urldef\tempurl%
\url{https://doi.org/10.1177/001872088702900304}
\showDOI{\tempurl}
\newblock
\shownote{Publisher: SAGE Publications Inc}.


\bibitem[{American Printing House for the Blind, Inc.}(1998)]%
        {americanprintinghousefortheblindinc.GoodTactileGraphic1998}
\bibfield{author}{\bibinfo{person}{{American Printing House for the Blind,
  Inc.}}} \bibinfo{year}{1998}\natexlab{}.
\newblock \bibinfo{title}{The {Good} {Tactile} {Graphic}}.
\newblock
\newblock
\urldef\tempurl%
\url{https://sites.aph.org/files/manuals/7-30006-00.pdf}
\showURL{%
\tempurl}


\bibitem[Anthony and Adams(2005)]%
        {anthonyDefinitionRoleTVI2005}
\bibfield{author}{\bibinfo{person}{Tanni Anthony} {and}
  \bibinfo{person}{Barbara Adams}.} \bibinfo{year}{2005}\natexlab{}.
\newblock \bibinfo{title}{The {Definition} and the {Role} of a {TVI}}.
\newblock
\newblock


\bibitem[{Ather Sharif}(2024)]%
        {athersharifAmDisabledNot2024}
\bibfield{author}{\bibinfo{person}{{Ather Sharif}}.}
  \bibinfo{year}{2024}\natexlab{}.
\newblock \bibinfo{title}{I am disabled but not â€˜impairedâ€™}.
\newblock
\newblock
\urldef\tempurl%
\url{https://interactions.acm.org/blog/view/i-am-disabled-but-not-impaired}
\showURL{%
\tempurl}


\bibitem[Baker et~al\mbox{.}(2016)]%
        {bakerTactileGraphicsVoice2016a}
\bibfield{author}{\bibinfo{person}{Catherine~M. Baker},
  \bibinfo{person}{Lauren~R. Milne}, \bibinfo{person}{Ryan Drapeau},
  \bibinfo{person}{Jeffrey Scofield}, \bibinfo{person}{Cynthia~L. Bennett},
  {and} \bibinfo{person}{Richard~E. Ladner}.} \bibinfo{year}{2016}\natexlab{}.
\newblock \showarticletitle{Tactile {Graphics} with a {Voice}}.
\newblock \bibinfo{journal}{\emph{ACM Transactions on Accessible Computing}}
  \bibinfo{volume}{8}, \bibinfo{number}{1} (\bibinfo{date}{Jan.}
  \bibinfo{year}{2016}), \bibinfo{pages}{3:1--3:22}.
\newblock
\showISSN{1936-7228}
\urldef\tempurl%
\url{https://doi.org/10.1145/2854005}
\showDOI{\tempurl}


\bibitem[Barth(1982)]%
        {barthTactileGraphicsGuidebook1982}
\bibfield{author}{\bibinfo{person}{John~L. Barth}.}
  \bibinfo{year}{1982}\natexlab{}.
\newblock \bibinfo{booktitle}{\emph{Tactile {Graphics} {Guidebook}}}.
\newblock \bibinfo{publisher}{American Printing House for the Blind},
  \bibinfo{address}{Louisville, KY}.
\newblock


\bibitem[Barth(1984)]%
        {doi:10.1177/001872088402600106}
\bibfield{author}{\bibinfo{person}{John~L. Barth}.}
  \bibinfo{year}{1984}\natexlab{}.
\newblock \showarticletitle{Incised grids: {Enhancing} the readability of
  tangible graphs for the blind}.
\newblock \bibinfo{journal}{\emph{Human Factors}} \bibinfo{volume}{26},
  \bibinfo{number}{1} (\bibinfo{year}{1984}), \bibinfo{pages}{61--70}.
\newblock
\urldef\tempurl%
\url{https://doi.org/10.1177/001872088402600106}
\showDOI{\tempurl}
\newblock
\shownote{tex.eprint: https://doi.org/10.1177/001872088402600106}.


\bibitem[{Braille Authority of North America}(2022)]%
        {brailleauthorityofnorthamericaGuidelinesStandardsTactile2022}
\bibfield{author}{\bibinfo{person}{{Braille Authority of North America}}.}
  \bibinfo{year}{2022}\natexlab{}.
\newblock \bibinfo{title}{Guidelines and {Standards} for {Tactile} {Graphics}
  {\textbar} {Braille} {Authority} of {North} {America}}.
\newblock
\newblock
\urldef\tempurl%
\url{https://www.brailleauthority.org/guidelines-and-standards-tactile-graphics}
\showURL{%
\tempurl}


\bibitem[Brauner(2024)]%
        {braunerTeachingNumberLine2024}
\bibfield{author}{\bibinfo{person}{Diane Brauner}.}
  \bibinfo{year}{2024}\natexlab{}.
\newblock \bibinfo{title}{Teaching number line math skills: {Part} 2}.
\newblock
\newblock
\urldef\tempurl%
\url{https://www.perkins.org/resource/teaching-number-line-math-skills-part-2/}
\showURL{%
\tempurl}


\bibitem[{Carmen Willings}(2020)]%
        {carmenwillingsTeacherStudentsVisual2020}
\bibfield{author}{\bibinfo{person}{{Carmen Willings}}.}
  \bibinfo{year}{2020}\natexlab{}.
\newblock \bibinfo{title}{Teacher of {Students} with {Visual} {Impairments}}.
\newblock
\newblock
\urldef\tempurl%
\url{https://www.teachingvisuallyimpaired.com/teacher-of-students-with-visual-impairments.html}
\showURL{%
\tempurl}


\bibitem[Challis and Edwards(2001)]%
        {challisDesignPrinciplesTactile2001}
\bibfield{author}{\bibinfo{person}{Ben~P. Challis} {and}
  \bibinfo{person}{Alistair~D.N. Edwards}.} \bibinfo{year}{2001}\natexlab{}.
\newblock \showarticletitle{Design principles for tactile interaction}. In
  \bibinfo{booktitle}{\emph{Haptic {Human}-{Computer} {Interaction}}}
  \emph{(\bibinfo{series}{Lecture {Notes} in {Computer} {Science}})},
  \bibfield{editor}{\bibinfo{person}{Stephen Brewster} {and}
  \bibinfo{person}{Roderick Murray-Smith}} (Eds.).
  \bibinfo{publisher}{Springer}, \bibinfo{address}{Berlin, Heidelberg},
  \bibinfo{pages}{17--24}.
\newblock
\showISBNx{978-3-540-44589-0}
\urldef\tempurl%
\url{https://doi.org/10.1007/3-540-44589-7_2}
\showDOI{\tempurl}


\bibitem[Chundury et~al\mbox{.}(2023)]%
        {chunduryTactualPlotSpatializingData2023}
\bibfield{author}{\bibinfo{person}{Pramod Chundury}, \bibinfo{person}{Yasmin
  Reyazuddin}, \bibinfo{person}{J.~Bern Jordan}, \bibinfo{person}{Jonathan
  Lazar}, {and} \bibinfo{person}{Niklas Elmqvist}.}
  \bibinfo{year}{2023}\natexlab{}.
\newblock \showarticletitle{{TactualPlot}: {Spatializing} {Data} as {Sound}
  using {Sensory} {Substitution} for {Touchscreen} {Accessibility}}.
\newblock \bibinfo{journal}{\emph{IEEE Transactions on Visualization and
  Computer Graphics}} \bibinfo{volume}{30}, \bibinfo{number}{1}
  (\bibinfo{year}{2023}), \bibinfo{pages}{1--11}.
\newblock
\showISSN{1077-2626, 1941-0506, 2160-9306}
\urldef\tempurl%
\url{https://doi.org/10.1109/TVCG.2023.3326937}
\showDOI{\tempurl}


\bibitem[Crombie et~al\mbox{.}(2004)]%
        {crombieBiggerPictureAutomated2004}
\bibfield{author}{\bibinfo{person}{David Crombie}, \bibinfo{person}{Roger
  Lenoir}, \bibinfo{person}{Neil McKenzie}, {and} \bibinfo{person}{George
  Ioannidis}.} \bibinfo{year}{2004}\natexlab{}.
\newblock \showarticletitle{The {Bigger} {Picture}: {Automated} {Production}
  {Tools} for {Tactile} {Graphics}}. In \bibinfo{booktitle}{\emph{Computers
  {Helping} {People} with {Special} {Needs}}},
  \bibfield{editor}{\bibinfo{person}{Klaus Miesenberger},
  \bibinfo{person}{Joachim Klaus}, \bibinfo{person}{Wolfgang~L. Zagler}, {and}
  \bibinfo{person}{Dominique Burger}} (Eds.). \bibinfo{publisher}{Springer},
  \bibinfo{address}{Berlin, Heidelberg}, \bibinfo{pages}{713--720}.
\newblock
\showISBNx{978-3-540-27817-7}
\urldef\tempurl%
\url{https://doi.org/10.1007/978-3-540-27817-7_106}
\showDOI{\tempurl}


\bibitem[{Duxbury Systems}(2024)]%
        {duxburysystemsQuickTacFreeSoftware2024}
\bibfield{author}{\bibinfo{person}{{Duxbury Systems}}.}
  \bibinfo{year}{2024}\natexlab{}.
\newblock \bibinfo{title}{{QuickTac} free software}.
\newblock
\newblock
\urldef\tempurl%
\url{https://www.duxburysystems.com/quicktac.asp}
\showURL{%
\tempurl}


\bibitem[Edman(1992)]%
        {edmanTactileGraphics1992}
\bibfield{author}{\bibinfo{person}{Polly~K Edman}.}
  \bibinfo{year}{1992}\natexlab{}.
\newblock \bibinfo{booktitle}{\emph{Tactile {Graphics}}}.
  Vol.~\bibinfo{volume}{15}.
\newblock \bibinfo{publisher}{American Foundation for the Blind}.
\newblock


\bibitem[Engel et~al\mbox{.}(2019)]%
        {engelSVGPlottAccessibleTool2019}
\bibfield{author}{\bibinfo{person}{Christin Engel},
  \bibinfo{person}{Emma~Franziska MÃ¼ller}, {and} \bibinfo{person}{Gerhard
  Weber}.} \bibinfo{year}{2019}\natexlab{}.
\newblock \showarticletitle{{SVGPlott}: an accessible tool to generate highly
  adaptable, accessible audio-tactile charts for and from blind and visually
  impaired people}. In \bibinfo{booktitle}{\emph{Proceedings of the 12th {ACM}
  {International} {Conference} on {PErvasive} {Technologies} {Related} to
  {Assistive} {Environments}}} \emph{(\bibinfo{series}{{PETRA} '19})}.
  \bibinfo{publisher}{Association for Computing Machinery},
  \bibinfo{address}{New York, NY, USA}, \bibinfo{pages}{186--195}.
\newblock
\showISBNx{978-1-4503-6232-0}
\urldef\tempurl%
\url{https://doi.org/10.1145/3316782.3316793}
\showDOI{\tempurl}


\bibitem[Engel et~al\mbox{.}(2021)]%
        {engelTactileHeatmapsNovel2021}
\bibfield{author}{\bibinfo{person}{Christin Engel},
  \bibinfo{person}{Emma~Franziska MÃ¼ller}, {and} \bibinfo{person}{Gerhard
  Weber}.} \bibinfo{year}{2021}\natexlab{}.
\newblock \showarticletitle{Tactile {Heatmaps}: {A} {Novel} {Visualisation}
  {Technique} for {Data} {Analysis} with {Tactile} {Charts}}. In
  \bibinfo{booktitle}{\emph{The 14th {PErvasive} {Technologies} {Related} to
  {Assistive} {Environments} {Conference}}} \emph{(\bibinfo{series}{{PETRA}
  2021})}. \bibinfo{publisher}{Association for Computing Machinery},
  \bibinfo{address}{New York, NY, USA}, \bibinfo{pages}{16--25}.
\newblock
\showISBNx{978-1-4503-8792-7}
\urldef\tempurl%
\url{https://doi.org/10.1145/3453892.3458045}
\showDOI{\tempurl}


\bibitem[Engel and Weber(2017a)]%
        {engelAnalysisTactileChart2017}
\bibfield{author}{\bibinfo{person}{Christin Engel} {and}
  \bibinfo{person}{Gerhard Weber}.} \bibinfo{year}{2017}\natexlab{a}.
\newblock \showarticletitle{Analysis of {Tactile} {Chart} {Design}}. In
  \bibinfo{booktitle}{\emph{Proceedings of the 10th {International}
  {Conference} on {PErvasive} {Technologies} {Related} to {Assistive}
  {Environments}}} \emph{(\bibinfo{series}{{PETRA} '17})}.
  \bibinfo{publisher}{Association for Computing Machinery},
  \bibinfo{address}{New York, NY, USA}, \bibinfo{pages}{197--200}.
\newblock
\showISBNx{978-1-4503-5227-7}
\urldef\tempurl%
\url{https://doi.org/10.1145/3056540.3064955}
\showDOI{\tempurl}


\bibitem[Engel and Weber(2017b)]%
        {engelImproveAccessibilityTactile2017}
\bibfield{author}{\bibinfo{person}{Christin Engel} {and}
  \bibinfo{person}{Gerhard Weber}.} \bibinfo{year}{2017}\natexlab{b}.
\newblock \showarticletitle{Improve the {Accessibility} of {Tactile} {Charts}}.
  In \bibinfo{booktitle}{\emph{Human-{Computer} {Interaction} - {INTERACT}
  2017}} \emph{(\bibinfo{series}{Lecture {Notes} in {Computer} {Science}})},
  \bibfield{editor}{\bibinfo{person}{Regina Bernhaupt}, \bibinfo{person}{Girish
  Dalvi}, \bibinfo{person}{Anirudha Joshi}, \bibinfo{person}{Devanuj
  K.~Balkrishan}, \bibinfo{person}{Jacki O'Neill}, {and} \bibinfo{person}{Marco
  Winckler}} (Eds.). \bibinfo{publisher}{Springer International Publishing},
  \bibinfo{address}{Cham}, \bibinfo{pages}{187--195}.
\newblock
\showISBNx{978-3-319-67744-6}
\urldef\tempurl%
\url{https://doi.org/10.1007/978-3-319-67744-6_12}
\showDOI{\tempurl}


\bibitem[Engel and Weber(2019)]%
        {engelUserStudyDetailed2019}
\bibfield{author}{\bibinfo{person}{Christin Engel} {and}
  \bibinfo{person}{Gerhard Weber}.} \bibinfo{year}{2019}\natexlab{}.
\newblock \bibinfo{title}{User {Study}: {A} {Detailed} {View} on the
  {Effectiveness} and {Design} of {Tactile} {Charts} {\textbar}
  {SpringerLink}}.
\newblock
\newblock
\urldef\tempurl%
\url{https://link-springer-com.libproxy.mit.edu/chapter/10.1007/978-3-030-29381-9_5}
\showURL{%
\tempurl}


\bibitem[Garcia-Mejia(2024)]%
        {garcia-mejiaCentralRoleTeacher2024}
\bibfield{author}{\bibinfo{person}{Diana Garcia-Mejia}.}
  \bibinfo{year}{2024}\natexlab{}.
\newblock \bibinfo{title}{The {Central} {Role} of the {Teacher} of {Students}
  with {Visual} {Impairments}}.
\newblock
\newblock
\urldef\tempurl%
\url{https://aphconnectcenter.org/familyconnect/education/iep-individualized-education-program-3-years-to-22-years-old/central-role-of-the-tvi/}
\showURL{%
\tempurl}


\bibitem[Goncu and Marriott(2008)]%
        {goncuTactileChartGeneration2008}
\bibfield{author}{\bibinfo{person}{Cagatay Goncu} {and} \bibinfo{person}{Kim
  Marriott}.} \bibinfo{year}{2008}\natexlab{}.
\newblock \showarticletitle{Tactile chart generation tool}. In
  \bibinfo{booktitle}{\emph{Proceedings of the 10th international {ACM}
  {SIGACCESS} conference on {Computers} and accessibility}}
  \emph{(\bibinfo{series}{Assets '08})}. \bibinfo{publisher}{Association for
  Computing Machinery}, \bibinfo{address}{New York, NY, USA},
  \bibinfo{pages}{255--256}.
\newblock
\showISBNx{978-1-59593-976-0}
\urldef\tempurl%
\url{https://doi.org/10.1145/1414471.1414525}
\showDOI{\tempurl}


\bibitem[Goncu et~al\mbox{.}(2010a)]%
        {goncuTactileDiagramsWorth2010}
\bibfield{author}{\bibinfo{person}{Cagatay Goncu}, \bibinfo{person}{Kim
  Marriott}, {and} \bibinfo{person}{Frances Aldrich}.}
  \bibinfo{year}{2010}\natexlab{a}.
\newblock \showarticletitle{Tactile {Diagrams}: {Worth} {Ten} {Thousand}
  {Words}?}. In \bibinfo{booktitle}{\emph{Diagrammatic {Representation} and
  {Inference}}} \emph{(\bibinfo{series}{Lecture {Notes} in {Computer}
  {Science}})}, \bibfield{editor}{\bibinfo{person}{Ashok~K. Goel},
  \bibinfo{person}{Mateja Jamnik}, {and} \bibinfo{person}{N.~Hari Narayanan}}
  (Eds.). \bibinfo{publisher}{Springer}, \bibinfo{address}{Berlin, Heidelberg},
  \bibinfo{pages}{257--263}.
\newblock
\showISBNx{978-3-642-14600-8}
\urldef\tempurl%
\url{https://doi.org/10.1007/978-3-642-14600-8_25}
\showDOI{\tempurl}


\bibitem[Goncu et~al\mbox{.}(2010b)]%
        {goncuUsabilityAccessibleBar2010}
\bibfield{author}{\bibinfo{person}{Cagatay Goncu}, \bibinfo{person}{Kim
  Marriott}, {and} \bibinfo{person}{John Hurst}.}
  \bibinfo{year}{2010}\natexlab{b}.
\newblock \showarticletitle{Usability of {Accessible} {Bar} {Charts}}.
\newblock In \bibinfo{booktitle}{\emph{Diagrammatic {Representation} and
  {Inference}}}, \bibfield{editor}{\bibinfo{person}{Ashok~K. Goel},
  \bibinfo{person}{Mateja Jamnik}, {and} \bibinfo{person}{N.~Hari Narayanan}}
  (Eds.). Vol.~\bibinfo{volume}{6170}. \bibinfo{publisher}{Springer Berlin
  Heidelberg}, \bibinfo{address}{Berlin, Heidelberg},
  \bibinfo{pages}{167--181}.
\newblock
\showISBNx{978-3-642-14599-5 978-3-642-14600-8}
\urldef\tempurl%
\url{https://doi.org/10.1007/978-3-642-14600-8_17}
\showDOI{\tempurl}
\newblock
\shownote{Series Title: Lecture Notes in Computer Science}.


\bibitem[Gonzalez et~al\mbox{.}(2019)]%
        {gonzalezTactiledMoreBetter2019}
\bibfield{author}{\bibinfo{person}{Ricardo Gonzalez}, \bibinfo{person}{Carlos
  Gonzalez}, {and} \bibinfo{person}{John~A. Guerra-Gomez}.}
  \bibinfo{year}{2019}\natexlab{}.
\newblock \showarticletitle{Tactiled: {Towards} {More} and {Better} {Tactile}
  {Graphics} {Using} {Machine} {Learning}}. In
  \bibinfo{booktitle}{\emph{Proceedings of the 21st {International} {ACM}
  {SIGACCESS} {Conference} on {Computers} and {Accessibility}}}
  \emph{(\bibinfo{series}{{ASSETS} '19})}. \bibinfo{publisher}{Association for
  Computing Machinery}, \bibinfo{address}{New York, NY, USA},
  \bibinfo{pages}{530--532}.
\newblock
\showISBNx{978-1-4503-6676-2}
\urldef\tempurl%
\url{https://doi.org/10.1145/3308561.3354613}
\showDOI{\tempurl}


\bibitem[Hanson et~al\mbox{.}(2015)]%
        {hansonWritingAccessibility2015}
\bibfield{author}{\bibinfo{person}{Vicki~L. Hanson}, \bibinfo{person}{Anna
  Cavender}, {and} \bibinfo{person}{Shari Trewin}.}
  \bibinfo{year}{2015}\natexlab{}.
\newblock \showarticletitle{Writing about accessibility}.
\newblock \bibinfo{journal}{\emph{Interactions}} \bibinfo{volume}{22},
  \bibinfo{number}{6} (\bibinfo{date}{Oct.} \bibinfo{year}{2015}),
  \bibinfo{pages}{62--65}.
\newblock
\showISSN{1072-5520, 1558-3449}
\urldef\tempurl%
\url{https://doi.org/10.1145/2828432}
\showDOI{\tempurl}


\bibitem[Hasty(2024)]%
        {hastyTeachingTactileGraphics2024}
\bibfield{author}{\bibinfo{person}{Lucia Hasty}.}
  \bibinfo{year}{2024}\natexlab{}.
\newblock \bibinfo{title}{Teaching {Tactile} {Graphics}}.
\newblock
\newblock
\urldef\tempurl%
\url{https://www.perkins.org/resource/teaching-tactile-graphics/}
\showURL{%
\tempurl}


\bibitem[Holloway et~al\mbox{.}(2024)]%
        {hollowayRefreshableTactileDisplays2024}
\bibfield{author}{\bibinfo{person}{Leona Holloway}, \bibinfo{person}{Peter
  Cracknell}, \bibinfo{person}{Kate Stephens}, \bibinfo{person}{Melissa
  Fanshawe}, \bibinfo{person}{Samuel Reinders}, \bibinfo{person}{Kim Marriott},
  {and} \bibinfo{person}{Matthew Butler}.} \bibinfo{year}{2024}\natexlab{}.
\newblock \bibinfo{title}{Refreshable {Tactile} {Displays} for {Accessible}
  {Data} {Visualisation}}.
\newblock
\newblock
\urldef\tempurl%
\url{https://doi.org/10.48550/arXiv.2401.15836}
\showDOI{\tempurl}
\newblock
\shownote{arXiv:2401.15836 [cs]}.


\bibitem[HospitÃ¡l(2024)]%
        {hospitalTipsReadingTactile2024}
\bibfield{author}{\bibinfo{person}{Laura HospitÃ¡l}.}
  \bibinfo{year}{2024}\natexlab{}.
\newblock \bibinfo{title}{Tips for {Reading} {Tactile} {Graphics} in {Science}
  with a {Focus} on {State} {Assessment}}.
\newblock
\newblock
\urldef\tempurl%
\url{https://www.perkins.org/resource/tips-reading-tactile-graphics-science-focus-state-assessment/}
\showURL{%
\tempurl}


\bibitem[{International Council on English Braille (ICEB)}(2024)]%
        {internationalcouncilonenglishbrailleicebUnifiedEnglishBraille2024}
\bibfield{author}{\bibinfo{person}{{International Council on English Braille
  (ICEB)}}.} \bibinfo{year}{2024}\natexlab{}.
\newblock \bibinfo{title}{Unified {English} {Braille} ({UEB})}.
\newblock
\newblock
\urldef\tempurl%
\url{https://www.iceb.org/ueb.html}
\showURL{%
\tempurl}


\bibitem[Jansen et~al\mbox{.}(2015)]%
        {jansenOpportunitiesChallengesData2015}
\bibfield{author}{\bibinfo{person}{Yvonne Jansen}, \bibinfo{person}{Pierre
  Dragicevic}, \bibinfo{person}{Petra Isenberg}, \bibinfo{person}{Jason
  Alexander}, \bibinfo{person}{Abhijit Karnik}, \bibinfo{person}{Johan Kildal},
  \bibinfo{person}{Sriram Subramanian}, {and} \bibinfo{person}{Kasper
  HornbÃ¦k}.} \bibinfo{year}{2015}\natexlab{}.
\newblock \showarticletitle{Opportunities and {Challenges} for {Data}
  {Physicalization}}. In \bibinfo{booktitle}{\emph{Proceedings of the 33rd
  {Annual} {ACM} {Conference} on {Human} {Factors} in {Computing} {Systems}}}
  \emph{(\bibinfo{series}{{CHI} '15})}. \bibinfo{publisher}{Association for
  Computing Machinery}, \bibinfo{address}{New York, NY, USA},
  \bibinfo{pages}{3227--3236}.
\newblock
\showISBNx{978-1-4503-3145-6}
\urldef\tempurl%
\url{https://doi.org/10.1145/2702123.2702180}
\showDOI{\tempurl}


\bibitem[Jayant et~al\mbox{.}(2007)]%
        {jayantAutomatedTactileGraphics2007}
\bibfield{author}{\bibinfo{person}{Chandrika Jayant}, \bibinfo{person}{Matt
  Renzelmann}, \bibinfo{person}{Dana Wen}, \bibinfo{person}{Satria Krisnandi},
  \bibinfo{person}{Richard Ladner}, {and} \bibinfo{person}{Dan Comden}.}
  \bibinfo{year}{2007}\natexlab{}.
\newblock \showarticletitle{Automated tactile graphics translation: in the
  field}. In \bibinfo{booktitle}{\emph{Proceedings of the 9th international
  {ACM} {SIGACCESS} conference on {Computers} and accessibility}}
  \emph{(\bibinfo{series}{Assets '07})}. \bibinfo{publisher}{Association for
  Computing Machinery}, \bibinfo{address}{New York, NY, USA},
  \bibinfo{pages}{75--82}.
\newblock
\showISBNx{978-1-59593-573-1}
\urldef\tempurl%
\url{https://doi.org/10.1145/1296843.1296858}
\showDOI{\tempurl}


\bibitem[Krufka et~al\mbox{.}(2007)]%
        {krufkaVisualTactileConversion2007a}
\bibfield{author}{\bibinfo{person}{Stephen~E. Krufka},
  \bibinfo{person}{Kenneth~E. Barner}, {and} \bibinfo{person}{Tuncer~Can
  Aysal}.} \bibinfo{year}{2007}\natexlab{}.
\newblock \showarticletitle{Visual to {Tactile} {Conversion} of {Vector}
  {Graphics}}.
\newblock \bibinfo{journal}{\emph{IEEE Transactions on Neural Systems and
  Rehabilitation Engineering}} \bibinfo{volume}{15}, \bibinfo{number}{2}
  (\bibinfo{date}{June} \bibinfo{year}{2007}), \bibinfo{pages}{310--321}.
\newblock
\showISSN{1558-0210}
\urldef\tempurl%
\url{https://doi.org/10.1109/TNSRE.2007.897029}
\showDOI{\tempurl}
\newblock
\shownote{Conference Name: IEEE Transactions on Neural Systems and
  Rehabilitation Engineering}.


\bibitem[Lederman and Campbell(1982)]%
        {ledermanTangibleGraphsBlind1982}
\bibfield{author}{\bibinfo{person}{Susan~J. Lederman} {and}
  \bibinfo{person}{Jamie~I. Campbell}.} \bibinfo{year}{1982}\natexlab{}.
\newblock \showarticletitle{Tangible {Graphs} for the {Blind}}.
\newblock \bibinfo{journal}{\emph{Human Factors}} \bibinfo{volume}{24},
  \bibinfo{number}{1} (\bibinfo{date}{Feb.} \bibinfo{year}{1982}),
  \bibinfo{pages}{85--100}.
\newblock
\showISSN{0018-7208}
\urldef\tempurl%
\url{https://doi.org/10.1177/001872088202400109}
\showDOI{\tempurl}
\newblock
\shownote{Publisher: SAGE Publications Inc}.


\bibitem[{Liblouis}(2024)]%
        {liblouisLiblouisOpensourceBraille2024}
\bibfield{author}{\bibinfo{person}{{Liblouis}}.}
  \bibinfo{year}{2024}\natexlab{}.
\newblock \bibinfo{title}{Liblouis - {An} open-source braille translator and
  back-translator.}
\newblock
\newblock
\urldef\tempurl%
\url{https://liblouis.io/}
\showURL{%
\tempurl}


\bibitem[Literacy(2014)]%
        {literacyCreatingLargePrint2014}
\bibfield{author}{\bibinfo{person}{Paths~to Literacy}.}
  \bibinfo{year}{2014}\natexlab{}.
\newblock \bibinfo{title}{Creating {Large} {Print} and {Tactile} {Graphs}}.
\newblock
\newblock
\urldef\tempurl%
\url{https://www.pathstoliteracy.org/creating-large-print-and-tactile-graphs/}
\showURL{%
\tempurl}


\bibitem[Mech et~al\mbox{.}(2014)]%
        {mechEdutactileToolRapid2014}
\bibfield{author}{\bibinfo{person}{Mrinal Mech}, \bibinfo{person}{Kunal
  Kwatra}, \bibinfo{person}{Supriya Das}, \bibinfo{person}{Piyush Chanana},
  \bibinfo{person}{Rohan Paul}, {and} \bibinfo{person}{M. Balakrishnan}.}
  \bibinfo{year}{2014}\natexlab{}.
\newblock \showarticletitle{Edutactile - {A} {Tool} for {Rapid} {Generation} of
  {Accurate} {Guideline}-{Compliant} {Tactile} {Graphics} for {Science} and
  {Mathematics}}. In \bibinfo{booktitle}{\emph{Computers {Helping} {People}
  with {Special} {Needs}}}, \bibfield{editor}{\bibinfo{person}{Klaus
  Miesenberger}, \bibinfo{person}{Deborah Fels}, \bibinfo{person}{Dominique
  Archambault}, \bibinfo{person}{Petr PeÅˆÃ¡z}, {and} \bibinfo{person}{Wolfgang
  Zagler}} (Eds.). \bibinfo{publisher}{Springer International Publishing},
  \bibinfo{address}{Cham}, \bibinfo{pages}{34--41}.
\newblock
\showISBNx{978-3-319-08599-9}
\urldef\tempurl%
\url{https://doi.org/10.1007/978-3-319-08599-9_6}
\showDOI{\tempurl}


\bibitem[Moured et~al\mbox{.}(2024)]%
        {10.1145/3640543.3645175}
\bibfield{author}{\bibinfo{person}{Omar Moured}, \bibinfo{person}{Morris
  Baumgarten-Egemole}, \bibinfo{person}{Karin MÃ¼ller}, \bibinfo{person}{Alina
  Roitberg}, \bibinfo{person}{Thorsten Schwarz}, {and} \bibinfo{person}{Rainer
  Stiefelhagen}.} \bibinfo{year}{2024}\natexlab{}.
\newblock \showarticletitle{{Chart4Blind}: {An} intelligent interface for chart
  accessibility conversion}. In \bibinfo{booktitle}{\emph{Proceedings of the
  29th international conference on intelligent user interfaces}}
  \emph{(\bibinfo{series}{Iui '24})}. \bibinfo{publisher}{Association for
  Computing Machinery}, \bibinfo{address}{New York, NY, USA},
  \bibinfo{pages}{504--514}.
\newblock
\showISBNx{9798400705083}
\urldef\tempurl%
\url{https://doi.org/10.1145/3640543.3645175}
\showDOI{\tempurl}
\newblock
\shownote{Number of pages: 11 Place: Greenville, SC, USA}.


\bibitem[Prescher et~al\mbox{.}(2014)]%
        {prescherProductionAccessibleTactile2014}
\bibfield{author}{\bibinfo{person}{Denise Prescher}, \bibinfo{person}{Jens
  Bornschein}, {and} \bibinfo{person}{Gerhard Weber}.}
  \bibinfo{year}{2014}\natexlab{}.
\newblock \showarticletitle{Production of {Accessible} {Tactile} {Graphics}}.
\newblock In \bibinfo{booktitle}{\emph{Computers {Helping} {People} with
  {Special} {Needs}}}, \bibfield{editor}{\bibinfo{person}{Klaus Miesenberger},
  \bibinfo{person}{Deborah Fels}, \bibinfo{person}{Dominique Archambault},
  \bibinfo{person}{Petr PeÅˆÃ¡z}, {and} \bibinfo{person}{Wolfgang Zagler}}
  (Eds.). Vol.~\bibinfo{volume}{8548}. \bibinfo{publisher}{Springer
  International Publishing}, \bibinfo{address}{Cham}, \bibinfo{pages}{26--33}.
\newblock
\showISBNx{978-3-319-08598-2 978-3-319-08599-9}
\urldef\tempurl%
\url{https://doi.org/10.1007/978-3-319-08599-9_5}
\showDOI{\tempurl}
\newblock
\shownote{Series Title: Lecture Notes in Computer Science}.


\bibitem[Prescher et~al\mbox{.}(2017)]%
        {prescherConsistencyTactilePattern2017}
\bibfield{author}{\bibinfo{person}{Denise Prescher}, \bibinfo{person}{Jens
  Bornschein}, {and} \bibinfo{person}{Gerhard Weber}.}
  \bibinfo{year}{2017}\natexlab{}.
\newblock \showarticletitle{Consistency of a {Tactile} {Pattern} {Set}}.
\newblock \bibinfo{journal}{\emph{ACM Transactions on Accessible Computing}}
  \bibinfo{volume}{10}, \bibinfo{number}{2} (\bibinfo{date}{April}
  \bibinfo{year}{2017}), \bibinfo{pages}{7:1--7:29}.
\newblock
\showISSN{1936-7228}
\urldef\tempurl%
\url{https://doi.org/10.1145/3053723}
\showDOI{\tempurl}


\bibitem[Race et~al\mbox{.}(2019)]%
        {raceDesigningTactileSchematics2019}
\bibfield{author}{\bibinfo{person}{Lauren Race}, \bibinfo{person}{Chancey
  Fleet}, \bibinfo{person}{Joshua~A. Miele}, \bibinfo{person}{Tom Igoe}, {and}
  \bibinfo{person}{Amy Hurst}.} \bibinfo{year}{2019}\natexlab{}.
\newblock \showarticletitle{Designing {Tactile} {Schematics}: {Improving}
  {Electronic} {Circuit} {Accessibility}}. In
  \bibinfo{booktitle}{\emph{Proceedings of the 21st {International} {ACM}
  {SIGACCESS} {Conference} on {Computers} and {Accessibility}}}
  \emph{(\bibinfo{series}{{ASSETS} '19})}. \bibinfo{publisher}{Association for
  Computing Machinery}, \bibinfo{address}{New York, NY, USA},
  \bibinfo{pages}{581--583}.
\newblock
\showISBNx{978-1-4503-6676-2}
\urldef\tempurl%
\url{https://doi.org/10.1145/3308561.3354610}
\showDOI{\tempurl}


\bibitem[Rosenberg(2024)]%
        {rosenbergTouchingNews2024}
\bibfield{author}{\bibinfo{person}{Naomi Rosenberg}.}
  \bibinfo{year}{2024}\natexlab{}.
\newblock \bibinfo{title}{Touching the {News}}.
\newblock
\newblock
\urldef\tempurl%
\url{https://lighthouse-sf.org/ttn/}
\showURL{%
\tempurl}


\bibitem[Rosling(2006)]%
        {roslingBestStatsYou2006}
\bibfield{author}{\bibinfo{person}{Hans Rosling}.}
  \bibinfo{year}{2006}\natexlab{}.
\newblock \bibinfo{title}{The best stats you've ever seen}.
\newblock
\newblock
\urldef\tempurl%
\url{https://www.ted.com/talks/hans_rosling_the_best_stats_you_ve_ever_seen}
\showURL{%
\tempurl}
\newblock
\shownote{https://www.ted.com/talks/hans\_rosling\_the\_best\_stats\_you\_ve\_ever\_seen}.


\bibitem[Satyanarayan et~al\mbox{.}(2017)]%
        {satyanarayanVegaLiteGrammarInteractive2017}
\bibfield{author}{\bibinfo{person}{Arvind Satyanarayan},
  \bibinfo{person}{Dominik Moritz}, \bibinfo{person}{Kanit Wongsuphasawat},
  {and} \bibinfo{person}{Jeffrey Heer}.} \bibinfo{year}{2017}\natexlab{}.
\newblock \showarticletitle{Vega-{Lite}: {A} {Grammar} of {Interactive}
  {Graphics}}.
\newblock \bibinfo{journal}{\emph{IEEE Transactions on Visualization and
  Computer Graphics}} \bibinfo{volume}{23}, \bibinfo{number}{1}
  (\bibinfo{date}{Jan.} \bibinfo{year}{2017}), \bibinfo{pages}{341--350}.
\newblock
\showISSN{1077-2626}
\urldef\tempurl%
\url{https://doi.org/10.1109/TVCG.2016.2599030}
\showDOI{\tempurl}


\bibitem[Seo et~al\mbox{.}(2024)]%
        {10.1145/3613904.3642730}
\bibfield{author}{\bibinfo{person}{JooYoung Seo}, \bibinfo{person}{Yilin Xia},
  \bibinfo{person}{Bongshin Lee}, \bibinfo{person}{Sean Mccurry}, {and}
  \bibinfo{person}{Yu~Jun Yam}.} \bibinfo{year}{2024}\natexlab{}.
\newblock \showarticletitle{{MAIDR}: {Making} statistical visualizations
  accessible with multimodal data representation}. In
  \bibinfo{booktitle}{\emph{Proceedings of the 2024 {CHI} conference on human
  factors in computing systems}} \emph{(\bibinfo{series}{Chi '24})}.
  \bibinfo{publisher}{Association for Computing Machinery},
  \bibinfo{address}{New York, NY, USA}, \bibinfo{pages}{1--22}.
\newblock
\showISBNx{9798400703300}
\urldef\tempurl%
\url{https://doi.org/10.1145/3613904.3642730}
\showDOI{\tempurl}
\newblock
\shownote{Number of pages: 22 Place: Honolulu, HI, USA tex.articleno: 211}.


\bibitem[Sharif et~al\mbox{.}(2021)]%
        {sharifUnderstandingScreenReaderUsers2021}
\bibfield{author}{\bibinfo{person}{Ather Sharif},
  \bibinfo{person}{Sanjana~Shivani Chintalapati}, \bibinfo{person}{Jacob~O.
  Wobbrock}, {and} \bibinfo{person}{Katharina Reinecke}.}
  \bibinfo{year}{2021}\natexlab{}.
\newblock \showarticletitle{Understanding {Screen}-{Reader} {Users}â€™
  {Experiences} with {Online} {Data} {Visualizations}}. In
  \bibinfo{booktitle}{\emph{Proceedings of the 23rd {International} {ACM}
  {SIGACCESS} {Conference} on {Computers} and {Accessibility}}}
  \emph{(\bibinfo{series}{{ASSETS} '21})}. \bibinfo{publisher}{Association for
  Computing Machinery}, \bibinfo{address}{New York, NY, USA},
  \bibinfo{pages}{1--16}.
\newblock
\showISBNx{978-1-4503-8306-6}
\urldef\tempurl%
\url{https://doi.org/10.1145/3441852.3471202}
\showDOI{\tempurl}


\bibitem[Sharif et~al\mbox{.}(2022)]%
        {sharifVoxLensMakingOnline2022}
\bibfield{author}{\bibinfo{person}{Ather Sharif}, \bibinfo{person}{Olivia~H.
  Wang}, \bibinfo{person}{Alida~T. Muongchan}, \bibinfo{person}{Katharina
  Reinecke}, {and} \bibinfo{person}{Jacob~O. Wobbrock}.}
  \bibinfo{year}{2022}\natexlab{}.
\newblock \showarticletitle{{VoxLens}: {Making} {Online} {Data}
  {Visualizations} {Accessible} with an {Interactive} {JavaScript}
  {Plug}-{In}}. In \bibinfo{booktitle}{\emph{Proceedings of the 2022 {CHI}
  {Conference} on {Human} {Factors} in {Computing} {Systems}}}
  \emph{(\bibinfo{series}{{CHI} '22})}. \bibinfo{publisher}{Association for
  Computing Machinery}, \bibinfo{address}{New York, NY, USA},
  \bibinfo{pages}{1--19}.
\newblock
\showISBNx{978-1-4503-9157-3}
\urldef\tempurl%
\url{https://doi.org/10.1145/3491102.3517431}
\showDOI{\tempurl}


\bibitem[Thompson et~al\mbox{.}(2023)]%
        {thompsonChartReaderAccessible2023}
\bibfield{author}{\bibinfo{person}{John~R Thompson}, \bibinfo{person}{Jesse~J
  Martinez}, \bibinfo{person}{Alper Sarikaya}, \bibinfo{person}{Edward
  Cutrell}, {and} \bibinfo{person}{Bongshin Lee}.}
  \bibinfo{year}{2023}\natexlab{}.
\newblock \showarticletitle{Chart {Reader}: {Accessible} {Visualization}
  {Experiences} {Designed} with {Screen} {Reader} {Users}}. In
  \bibinfo{booktitle}{\emph{Proceedings of the 2023 {CHI} {Conference} on
  {Human} {Factors} in {Computing} {Systems}}} \emph{(\bibinfo{series}{{CHI}
  '23})}. \bibinfo{publisher}{Association for Computing Machinery},
  \bibinfo{address}{New York, NY, USA}, \bibinfo{pages}{1--18}.
\newblock
\showISBNx{978-1-4503-9421-5}
\urldef\tempurl%
\url{https://doi.org/10.1145/3544548.3581186}
\showDOI{\tempurl}


\bibitem[University(2002)]%
        {purdueuniversityTactileDiagramManual2002}
\bibfield{author}{\bibinfo{person}{Purdue University}.}
  \bibinfo{year}{2002}\natexlab{}.
\newblock \bibinfo{title}{Tactile {Diagram} {Manual} (2002, {Print}
  {Version})}.
\newblock
\newblock
\urldef\tempurl%
\url{https://www.yumpu.com/en/document/view/25596604/tactile-diagram-manual-2002-print-version-purdue-university}
\showURL{%
\tempurl}


\bibitem[Watanabe et~al\mbox{.}(2016)]%
        {watanabeDevelopmentTactileGraph2016}
\bibfield{author}{\bibinfo{person}{Tetsuya Watanabe}, \bibinfo{person}{Kosuke
  Araki}, \bibinfo{person}{Toshimitsu Yamaguchi}, {and}
  \bibinfo{person}{Kazunori Minatani}.} \bibinfo{year}{2016}\natexlab{}.
\newblock \showarticletitle{Development of {Tactile} {Graph} {Generation} {Web}
  {Application} {Using} {R} {Statistics} {Software} {Environment}}.
\newblock \bibinfo{journal}{\emph{IEICE Transactions on Information and
  Systems}} \bibinfo{volume}{E99.D}, \bibinfo{number}{8}
  (\bibinfo{year}{2016}), \bibinfo{pages}{2151--2160}.
\newblock
\urldef\tempurl%
\url{https://doi.org/10.1587/transinf.2015EDP7405}
\showDOI{\tempurl}


\bibitem[Watanabe and Inaba(2018)]%
        {watanabeTexturesSuitableTactile2018}
\bibfield{author}{\bibinfo{person}{Tetsuya Watanabe} {and}
  \bibinfo{person}{Naoki Inaba}.} \bibinfo{year}{2018}\natexlab{}.
\newblock \showarticletitle{Textures {Suitable} for {Tactile} {Bar} {Charts} on
  {Capsule} {Paper}}.
\newblock \bibinfo{journal}{\emph{Transactions of the Virtual Reality Society
  of Japan}}  \bibinfo{volume}{23} (\bibinfo{year}{2018}),
  \bibinfo{pages}{13--20}.
\newblock


\bibitem[Watanabe and Mizukami(2018)]%
        {watanabeEffectivenessTactileScatter2018}
\bibfield{author}{\bibinfo{person}{Tetsuya Watanabe} {and}
  \bibinfo{person}{Hikaru Mizukami}.} \bibinfo{year}{2018}\natexlab{}.
\newblock \showarticletitle{Effectiveness of {Tactile} {Scatter} {Plots}:
  {Comparison} of {Non}-visual {Data} {Representations}}. In
  \bibinfo{booktitle}{\emph{Computers {Helping} {People} with {Special}
  {Needs}}} \emph{(\bibinfo{series}{Lecture {Notes} in {Computer} {Science}})},
  \bibfield{editor}{\bibinfo{person}{Klaus Miesenberger} {and}
  \bibinfo{person}{Georgios Kouroupetroglou}} (Eds.).
  \bibinfo{publisher}{Springer International Publishing},
  \bibinfo{address}{Cham}, \bibinfo{pages}{628--635}.
\newblock
\showISBNx{978-3-319-94277-3}
\urldef\tempurl%
\url{https://doi.org/10.1007/978-3-319-94277-3_97}
\showDOI{\tempurl}


\bibitem[Watanabe et~al\mbox{.}(2012)]%
        {watanabeDevelopmentSoftwareAutomatic2012}
\bibfield{author}{\bibinfo{person}{Tetsuya Watanabe},
  \bibinfo{person}{Toshimitsu Yamaguchi}, {and} \bibinfo{person}{Masaki
  Nakagawa}.} \bibinfo{year}{2012}\natexlab{}.
\newblock \showarticletitle{Development of {Software} for {Automatic}
  {Creation} of {Embossed} {Graphs}}.
\newblock In \bibinfo{booktitle}{\emph{Computers {Helping} {People} with
  {Special} {Needs}}}, \bibfield{editor}{\bibinfo{person}{David Hutchison},
  \bibinfo{person}{Takeo Kanade}, \bibinfo{person}{Josef Kittler},
  \bibinfo{person}{Jon~M. Kleinberg}, \bibinfo{person}{Friedemann Mattern},
  \bibinfo{person}{John~C. Mitchell}, \bibinfo{person}{Moni Naor},
  \bibinfo{person}{Oscar Nierstrasz}, \bibinfo{person}{C.~Pandu~Rangan},
  \bibinfo{person}{Bernhard Steffen}, \bibinfo{person}{Madhu Sudan},
  \bibinfo{person}{Demetri Terzopoulos}, \bibinfo{person}{Doug Tygar},
  \bibinfo{person}{Moshe~Y. Vardi}, \bibinfo{person}{Gerhard Weikum},
  \bibinfo{person}{Klaus Miesenberger}, \bibinfo{person}{Arthur Karshmer},
  \bibinfo{person}{Petr Penaz}, {and} \bibinfo{person}{Wolfgang Zagler}}
  (Eds.). Vol.~\bibinfo{volume}{7382}. \bibinfo{publisher}{Springer Berlin
  Heidelberg}, \bibinfo{address}{Berlin, Heidelberg},
  \bibinfo{pages}{174--181}.
\newblock
\showISBNx{978-3-642-31521-3 978-3-642-31522-0}
\urldef\tempurl%
\url{https://doi.org/10.1007/978-3-642-31522-0_25}
\showDOI{\tempurl}
\newblock
\shownote{Series Title: Lecture Notes in Computer Science}.


\bibitem[Way and Barner(1997)]%
        {wayAutomaticVisualTactile1997}
\bibfield{author}{\bibinfo{person}{T.P. Way} {and} \bibinfo{person}{K.E.
  Barner}.} \bibinfo{year}{1997}\natexlab{}.
\newblock \showarticletitle{Automatic visual to tactile translation. {I}.
  {Human} factors, access methods and image manipulation}.
\newblock \bibinfo{journal}{\emph{IEEE Transactions on Rehabilitation
  Engineering}} \bibinfo{volume}{5}, \bibinfo{number}{1} (\bibinfo{date}{March}
  \bibinfo{year}{1997}), \bibinfo{pages}{81--94}.
\newblock
\showISSN{1558-0024}
\urldef\tempurl%
\url{https://doi.org/10.1109/86.559353}
\showDOI{\tempurl}
\newblock
\shownote{Conference Name: IEEE Transactions on Rehabilitation Engineering}.


\bibitem[Yang et~al\mbox{.}(2020)]%
        {yangTactilePresentationNetwork2020}
\bibfield{author}{\bibinfo{person}{Yalong Yang}, \bibinfo{person}{Kim
  Marriott}, \bibinfo{person}{Matthew Butler}, \bibinfo{person}{Cagatay Goncu},
  {and} \bibinfo{person}{Leona Holloway}.} \bibinfo{year}{2020}\natexlab{}.
\newblock \showarticletitle{Tactile {Presentation} of {Network} {Data}: {Text},
  {Matrix} or {Diagram}?}. In \bibinfo{booktitle}{\emph{Proceedings of the 2020
  {CHI} {Conference} on {Human} {Factors} in {Computing} {Systems}}}
  \emph{(\bibinfo{series}{{CHI} '20})}. \bibinfo{publisher}{Association for
  Computing Machinery}, \bibinfo{address}{New York, NY, USA},
  \bibinfo{pages}{1--12}.
\newblock
\showISBNx{978-1-4503-6708-0}
\urldef\tempurl%
\url{https://doi.org/10.1145/3313831.3376367}
\showDOI{\tempurl}


\bibitem[Zong et~al\mbox{.}(2024)]%
        {zongUmweltAccessibleStructured2024b}
\bibfield{author}{\bibinfo{person}{Jonathan Zong}, \bibinfo{person}{Isabella
  Pedraza~Pineros}, \bibinfo{person}{Mengzhu~(Katie) Chen},
  \bibinfo{person}{Daniel Hajas}, {and} \bibinfo{person}{Arvind Satyanarayan}.}
  \bibinfo{year}{2024}\natexlab{}.
\newblock \showarticletitle{Umwelt: {Accessible} {Structured} {Editing} of
  {Multi}-{Modal} {Data} {Representations}}. In
  \bibinfo{booktitle}{\emph{Proceedings of the {CHI} {Conference} on {Human}
  {Factors} in {Computing} {Systems}}} \emph{(\bibinfo{series}{{CHI} '24})}.
  \bibinfo{publisher}{Association for Computing Machinery},
  \bibinfo{address}{New York, NY, USA}, \bibinfo{pages}{1--20}.
\newblock
\showISBNx{9798400703300}
\urldef\tempurl%
\url{https://doi.org/10.1145/3613904.3641996}
\showDOI{\tempurl}


\end{thebibliography}

\onecolumn
\clearpage
\appendix

\section{Appendix: Tactile Vega-Lite defaults}
\label{sec:default-values}

\begin{table*}[!ht]
\caption{Default configuration values for tactile charts in Tactile Vega-Lite, categorized by key design elements such as Braille, alignment, and tactual hierarchy. The rationale for the default values is primarily derived from established tactile design guidelines and insights gathered from our formative interviews about existing industry practices.}
\label{tab:default-values}
  \renewcommand{\arraystretch}{1.2} %
\begin{tabular}{p{3cm}p{2cm}p{4cm}p{5cm}}
\toprule
\textbf{Property} & \textbf{Default Value} & \textbf{Description} &
\textbf{Guidelines} \\ \midrule

\multicolumn{4}{l}{Braille} \\ \midrule
config.font & "Swell Braille" & Font used for text in the chart. & Swell Braille produces high-quality braille on both embossers and Swell Form machines. \\

titleFontSize, labelFontSize, subtitleFontSize (title, axis, legend) & 24 & Font size for default braille font Swell Braille & Each braille font has a fixed size for standard readability, as altering the size makes it unreadable. \\

axis.titleFontWeight, title.fontWeight, axis.labelFontWeight & "normal" & Braille fonts do not have different font weights. & Braille text must maintain uniform weight to preserve tactile clarity. \\

brailleTranslation & "en-ueb-g2.ctb" & Braille translation table used, including braille grade, language and braille code. &  This translation table converts English text into Grade 2 contracted braille, adhering to the UEB standard, which is widely adopted by braille readers in English-speaking countries.\\ \midrule

\multicolumn{4}{l}{Tactual Hierarchy} \\ \midrule

axis.gridWidth & 1 & Width of the gridlines. & Guideline 6.6.2.2: Lines Grid lines should be the least distinct lines on the graph. \\

axis.domainWidth & 2.5 & Width of the axis lines. & Guideline 6.6.2.2 Lines
The x-axis (horizontal) and y-axis (vertical) lines must be tactually distinct and stronger than the grid lines. \\

axis.tickSize & 26.5(px) & Default tick mark length. & Guideline minimum sizing rule. \\

axis.tickWidth & 2.5(px) & Line thickness of the axis ticks. & Same line width as the axis lines to ensure the same level of tactual hierarchy. \\

axis.gridColor, axis.domainColor, axis.tickColor & "black" & Color of the axis ticks. & The color black generally prints well on the embosser and swell form machine. \\

axis.staggerLabels & "auto" & Automatically stagger x axis label when label length exceeds threshold. & \\ \bottomrule
\end{tabular}
\end{table*}

\begin{table*}[!ht]
\caption{Default Values for Tactile Charts in Tactile Vega-Lite Cont}
\label{tab:default-values-2}
  \renewcommand{\arraystretch}{1.2} %
\begin{tabular}{p{3cm}p{2cm}p{4cm}p{5cm}}
\toprule
\textbf{Property} & \textbf{Default Value} & \textbf{Description} &
\textbf{Guidelines} \\ \midrule
\multicolumn{4}{l}{Positioning} \\ \midrule

legend.direction & "vertical" & Stacks legend entries vertically. & Vertical stacking supports easier tactile navigation for certain readers. \\

legend.orient & "top-left" & Determines the position of the legend in the chart. & Placing the legend in a predictable location improves usability for tactile users by minimizing search time. \\

axis.titleAngle, axis.labelAngle & 0 & Rotation angle for axis titles (0 means no rotation). & Based on tactile reading habits, users usually do not anticipate rotated angles and might be confused or take longer to read, as tactile reading relies on uniformity and consistency for efficient interpretation. \\ \midrule

\multicolumn{4}{l}{Alignment} \\ \midrule
title.align & "center" & Alignment of chart title and other elements. & Guideline 5.3.1: The most commonly used heading in a graphic is the centered heading. It is used for the title of a graphic. \\

axis.titleAlign & "left" & Left align title text. & Guideline 6.6.4.5: The heading label for the horizontal values should be placed below the values and should be left-justified with the first cell of the first horizontal value.\\ \bottomrule

\end{tabular}
\end{table*}

\begin{table*}[!ht]
\caption{Default Values for Tactile Charts in Tactile Vega-Lite Cont}
\label{tab:default-values-3}
  \renewcommand{\arraystretch}{1.2} %
\begin{tabular}{p{3cm}p{2cm}p{4cm}p{5cm}}
\toprule
\textbf{Property} & \textbf{Default Value} & \textbf{Description} &
\textbf{Guidelines} \\ \midrule
\multicolumn{4}{l}{Spacing and Size} \\ \midrule

title.offset & 50 & Add spacing between the title and chart area. & Guideline 5.3.1: Blank lines should be left before and after centered headings. \\

axis.titlePadding & 20 & Padding, in pixels, between the axis title and the axis. & Guideline 6.6.4.5: The heading label (axis title) for the horizontal values should be placed below the values. \\

axis.labelPadding & 20(px) & Padding, in pixels, between labels and axis ticks. & Guideline 6.6.4.5: On the horizontal axis, value should be spaced 1/8 inch from the tick mark or axis line.\\

axis.titleY & -10(px) & Y-coordinate offset for the axis title relative to the axis group. & Guideline minimum spacing rule of 1/8 inch. \\

legend.titlePadding & 20(px) & Padding, in pixels, between the legend title and the legend. & Guideline minimum spacing rule of 1/8 inch. \\

legend.offset & 20(px) & Padding between the bottom legend and the top of the chart. & Guideline minimum spacing rule of 1/8 inch.\\

legend.columnPadding & 20(px) & Padding between legend columns. & Guideline minimum spacing rule of 1/8 inch.\\

legend.rowPadding & 20(px) & Padding between legend rows. & Guideline minimum spacing rule of 1/8 inch.\\

config.padding &  \begin{tabular}[t]{@{}p{4cm}@{}}
    \{"top": 100, \\
    "bottom": 100, \\
    "left": 200, \\
    "right": 200\}
\end{tabular} & Padding around the chart to ensure elements are not cut off. & Guideline minimum spacing rule of 1/8 inch.\\

legend.symbolSize & 3000 & Size of legend symbols. & Guideline minimum sizing rule. \\ \bottomrule

\end{tabular}
\end{table*}

\end{document}